%% file: acl_latex.tex
\documentclass[11pt]{article}

\usepackage[]{acl}

\usepackage{times}
\usepackage{latexsym}

\usepackage[T1]{fontenc}

\usepackage[utf8]{inputenc}

\usepackage{microtype}

\usepackage{inconsolata}

\usepackage{graphicx}

\usepackage[most]{tcolorbox}

\input{preamble}

\title{MemRec: Collaborative Memory-Augmented\\ Agentic Recommender System}

\author{
    Weixin Chen$^{1,2}$ \quad
    Yuhan Zhao$^{1}$ \quad
    Jingyuan Huang$^{2}$ \quad
    Zihe Ye$^{2}$ \\
    \textbf{Clark Mingxuan Ju}$^{3\dagger}$ \quad 
    \textbf{Tong Zhao}$^{3\dagger}$ \quad 
    \textbf{Neil Shah}$^{3\dagger}$ \quad 
    \textbf{Li Chen}$^{1}$ \quad
    \textbf{Yongfeng Zhang}$^{2\ddagger}$ \\
    $^1$Hong Kong Baptist University \quad
    $^2$Rutgers University \quad
    $^3$Snap Inc. \\
    \texttt{\{cswxchen, csyhzhao, lichen\}@comp.hkbu.edu.hk} \\
    \texttt{\{chy.huang, zihe.ye, yongfeng.zhang\}@rutgers.edu} \\
    \texttt{\{mju, tong, nshah\}@snap.com}
}

\begin{document}
\maketitle

\renewcommand{\thefootnote}{\fnsymbol{footnote}}
\footnotetext[2]{Authors affiliated with Snap Inc. served in advisory roles only for this work.}
\footnotetext[3]{Corresponding author.}

\input{abstract}

\input{introduction}

\input{methodology}
\input{experiments}


\input{related_work}

\input{conclusion}
\clearpage
\input{limitations}

\input{acknowledgements}

\bibliography{custom}

\clearpage
\input{appendix}

\end{document}

%% file: preamble.tex
\usepackage{booktabs} 
\usepackage{listings} 
\usepackage{xcolor}   
\usepackage{amsmath}  
\usepackage{textcomp}  
\usepackage{enumitem}  
\usepackage{framed}   
\usepackage{colortbl}
\usepackage{subcaption}
\usepackage{xcolor}

\usepackage{amsfonts} 
\usepackage{algorithm}     
\usepackage{algorithmicx}  
\usepackage{algpseudocode} 

\definecolor{promptbg}{rgb}{0.97, 0.98, 1.0}   
\definecolor{rulebg}{rgb}{0.98, 1.0, 0.98}     
\definecolor{promptkeyword}{rgb}{0.0, 0.2, 0.7} 

\lstdefinestyle{promptstyle}{
    backgroundcolor=\color{promptbg},
    basicstyle=\ttfamily\small, 
    breaklines=true,
    columns=fullflexible,
    keepspaces=true,
    showstringspaces=false,
    frame=tb, 
    framerule=0pt,
    framexleftmargin=1em,
    framexrightmargin=1em,
    captionpos=b,
    moredelim=[is][\color{promptkeyword}]{@@}{@@} 
}

\lstdefinestyle{rulesstyle-plain}{
    backgroundcolor=\color{rulebg},
    basicstyle=\ttfamily\small,
    breaklines=true,
    columns=fullflexible,
    keepspaces=true,
    showstringspaces=false,
    frame=none,
    numbers=none, 
    xleftmargin=1em, 
    mathescape=true,
    literate={_}{{\_}}1 
}

\usepackage[most]{tcolorbox} 

\newtcolorbox{promptbox}[1][]{
  colback=gray!5!white,    
  colframe=gray!75!black,  
  title={\textbf{#1}},     
  fonttitle=\sffamily,     
  boxrule=0.8pt,           
  arc=2pt,                 
  left=5pt, right=5pt, top=5pt, bottom=5pt, 
  enhanced,                
  breakable                
}

\definecolor{promptkeyword}{rgb}{0.0, 0.2, 0.7} 
\definecolor{featurecolor}{rgb}{0.0, 0.3, 0.5} 
\definecolor{commentcolor}{rgb}{0.4, 0.4, 0.4} 

\newcommand{\feat}[1]{\textcolor{featurecolor}{\texttt{#1}}}
\newcommand{\rat}[1]{\textcolor{commentcolor}{\textit{#1}}}

\usepackage{titletoc}

\titlecontents{section}[0em] 
{\addvspace{0.4em}\bfseries} 
{\contentslabel{1.5em}} 
{} 
{\titlerule*[0.7pc]{.}\contentspage} 

\titlecontents{subsection}[2em] 
{\addvspace{0.1em}\small} 
{\contentslabel{2.0em}} 
{}
{\titlerule*[0.7pc]{.}\contentspage}

\usepackage{tabularx}

\usepackage{booktabs}

 \usepackage{multirow}

\definecolor{collabHighlight}{HTML}{1f77b4} 
\definecolor{userHighlight}{HTML}{ff7f0e}   
\newcommand{\collab}[1]{\textbf{\textcolor{collabHighlight}{#1}}} 
\newcommand{\user}[1]{\textbf{\textcolor{userHighlight}{#1}}} 

\usepackage{microtype}

\AtBeginDocument{
  \setlength{\abovedisplayskip}{4pt}
  \setlength{\belowdisplayskip}{4pt}
}

%% file: abstract.tex
\begin{abstract}


The evolution of recommender systems has shifted from traditional collaborative filtering to LLM-based agentic systems, which rely on semantic user and item memories to make predictions. However, existing agents maintain these memories in isolation. This overlooks crucial collaborative signals, such as user-item co-engagements and peer relationships across the community, which significantly limits their ability to uncover hidden preferences and accurately infer user needs, particularly for data-sparse users. To bridge this gap, we introduce \textit{collaborative memory}, a paradigm that connects isolated semantics to enable the sharing of relational insights. Yet, naively utilizing collaborative memory causes severe context overload and introduces noise to downstream LLMs, alongside prohibitive computational costs. To resolve this, we propose \textbf{\texttt{MemRec}}, a framework that architecturally decouples memory management from reasoning. \texttt{MemRec} introduces a dedicated, lightweight language model ($\text{LM}_{\text{Mem}}$) to efficiently manage and synthesize a dynamic collaborative memory graph in the background. It provides only distilled, high-signal contexts to a downstream, heavyweight large language model ($\text{LLM}_{\text{Rec}}$) for the final recommendation. Extensive experiments on four benchmarks demonstrate that \texttt{MemRec} achieves state-of-the-art performance. 

\noindent
\textit{\textbf{Code:}} \href{https://github.com/rutgerswiselab/memrec}{https://github.com/rutgerswiselab/memrec} 

\textit{\textbf{Homepage:}} \href{https://memrec.weixinchen.com}{https://memrec.weixinchen.com}

\end{abstract}

%% file: introduction.tex
\begin{figure}[t]
    \centering
    \subfloat[Traditional Agentic RS]{
        \includegraphics[width=0.45\columnwidth]{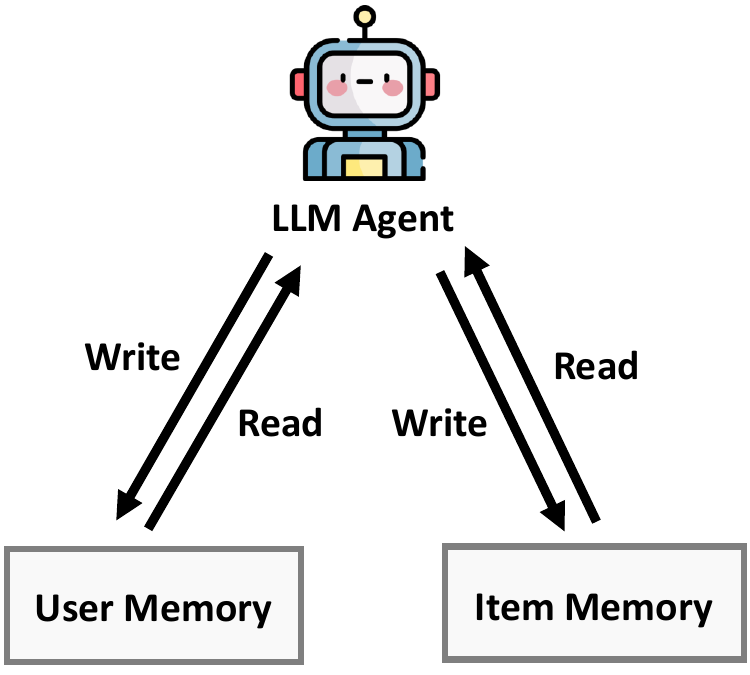} 
        \label{fig:memory_comparison_a}
    }
    \hfill
    \subfloat[MemRec Framework]{
        \includegraphics[width=0.46\columnwidth]{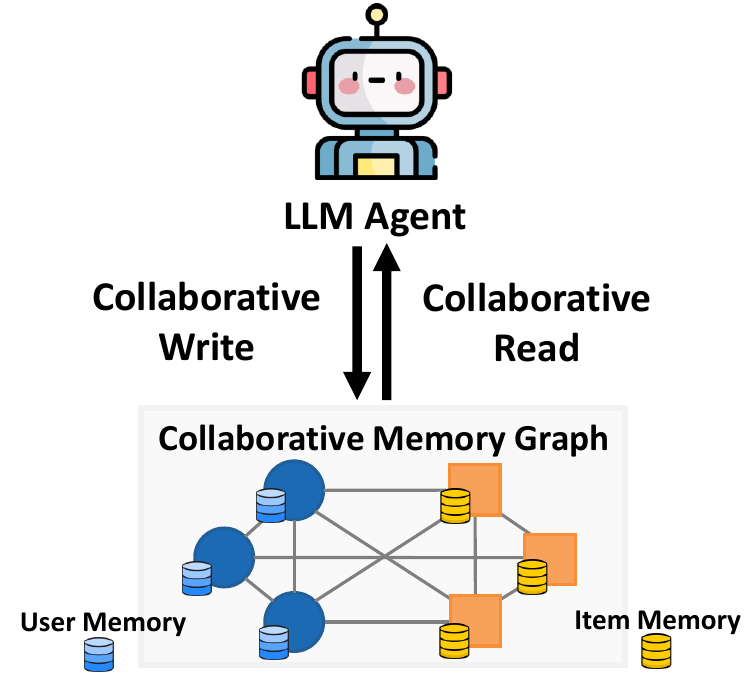} 
        \label{fig:memory_comparison_b}
    }
    \vspace{-2mm}
    \caption{
    \textbf{(a) Existing Agents} interact with user and item memories through separate, isolated read/write channels. \textbf{(b) MemRec} performs collaborative operations on a memory graph, enabling the flow of relational signals across peer users and related items to overcome data sparsity and transfer collaborative knowledge.
    }
    \label{fig:memory_comparison}
    \vspace{-4mm}
\end{figure}

\section{Introduction}
\label{sec:introduction}

Memory has long served as a foundational component in Recommender Systems (RS).  The field has evolved from capturing preferences through sparse rating matrices in conventional collaborative filtering era~\cite{sarwar2001item, koren2009matrix} to using dense latent embeddings in the deep learning era~\cite{covington2016deep, he2017neural}. Recently, the emergence of agentic RS, powered by Large Language Models (LLMs), has ushered in a new paradigm, i.e., semantic memory~\cite{wu2024survey, zhang2025survey}.

In agentic RS, memory is transformed into a semantic format, enabling LLMs to perform complex reasoning and use tools with natural language as the substrate~\cite{zhao2024let}. For an agent to be more than a stateless function, it must utilize this persistent memory to retain and evolve its user understanding through ongoing interactions~\cite{xi2025rise, park2023generative}. The evolution of memory mechanisms in this context can be delineated into three key milestones: (1) \textit{No explicit memory}, relying solely on the LLM's inherent knowledge~\cite{liu2023chatgpt, lyu2024llm}; (2) \textit{Static memory}, characterized by retrieving context from fixed storage~\cite{xu2025iagent, gao2023chat}; and recently (3) \textit{Dynamic, self-reflective memory}, where agents iteratively update their understanding over time~\cite{tang2025interactive,zhang2024agentcf}.


However, these approaches predominantly represent \textit{non-collaborative} paradigms. As illustrated in Figure \ref{fig:memory_comparison_a}, current agents typically reflect only on their siloed memories, such as a user's memory ($M_u$) containing a textual narrative of their past interactions, or an item's memory ($M_i$) initialized by its static semantic description~\cite{xu2025iagent, zhang2024agentcf}. This strictly local scope isolates them from the most potent signal in recommender systems: collaborative relationships. By ignoring the broader user-item graph, these agents fail to transfer interaction signals from warm to cold items through co-engagements, or to leverage the shared preferences of peer users to uncover hidden interests~\cite{wang2019neural, he2020lightgcn}.

A seemingly intuitive solution for bridging this gap is to inject raw collaborative neighborhoods directly into the agent's memory (e.g., naively concatenating a target user's context window with the textual interaction histories of dozens of similar peers, or expanding an item's memory with descriptions from all co-engaged items). However, this naive brute-force approach proves inadequate for at least two critical reasons:
\begin{itemize}[leftmargin=*, itemsep=1pt, topsep=3pt]
    \item \textit{Cognitive Overload.} While the agent may be able to access large quantities of neighbor memories, it struggles to effectively distill pertinent information from this abundance. The sheer volume of textual and structural signals increases difficulty for the reasoning agent to identify salient knowledge~\cite{liu2024lost}, as validated in \S\ref{sec:architectural_analysis}.
    \item \textit{Prohibitive Collaborative Updates.} A truly collaborative memory must evolve dynamically. Whenever a new interaction occurs, the updated knowledge should ideally propagate to all connected neighbors. Under a naive brute-force approach, keeping these neighborhood contexts synchronized necessitates redundant, independent LLM calls for every related user and item. This creates an intractable computational bottleneck during real-time serving, making continuous graph evolution prohibitively expensive.
\end{itemize}
Consequently, a core challenge emerges: \textit{How can we distill extensive collaborative knowledge into memory to empower the reasoning agent, while ensuring efficient evolution of the graph?}

To address these challenges, we introduce \textbf{\texttt{MemRec}} (Figure \ref{fig:memory_comparison_b}), a framework built upon architectural decoupling to shift from isolated to collaborative memory. By dedicating a separate Memory Manager (\textbf{$\text{LM}_{\text{Mem}}$}) to manage a dynamic graph and synthesize compact grounding, this architecture systematically resolves both cognitive overload and update bottlenecks.
Firstly, addressing cognitive overload during retrieval, our \textit{Collaborative Memory Retrieval} method overcomes the limitations of isolated memory paradigms. Instead of relying solely on siloed user or item memory, it leverages LLM-guided domain-adaptive rules to curates neighbor signals to synthesize a compact, high-utility collaborative memory.
Secondly, overcoming update bottlenecks, we develop an \textit{Asynchronous Collaborative Propagation} mechanism inspired by Label Propagation~\cite{zhu2002learning}. It efficiently batches self-reflection and neighbor updates into a single asynchronous operation and achieves constant-time ($O(1)$) interaction complexity, ensuring continuous graph evolution without incurring the computational penalties of redundant, independent updates.


Extensive evaluations on four benchmarks show that MemRec achieves state-of-the-art performance. Furthermore, our architectural analysis demonstrates MemRec's flexibility, establishing a new Pareto frontier that balances reasoning quality, computational cost, and deployment constraints, supporting diverse setups from cloud-native APIs to on-premise local models.
\vspace{-2mm}

%% file: methodology.tex
\begin{figure*}[t]

    \centering

    \includegraphics[width=1.00\textwidth]{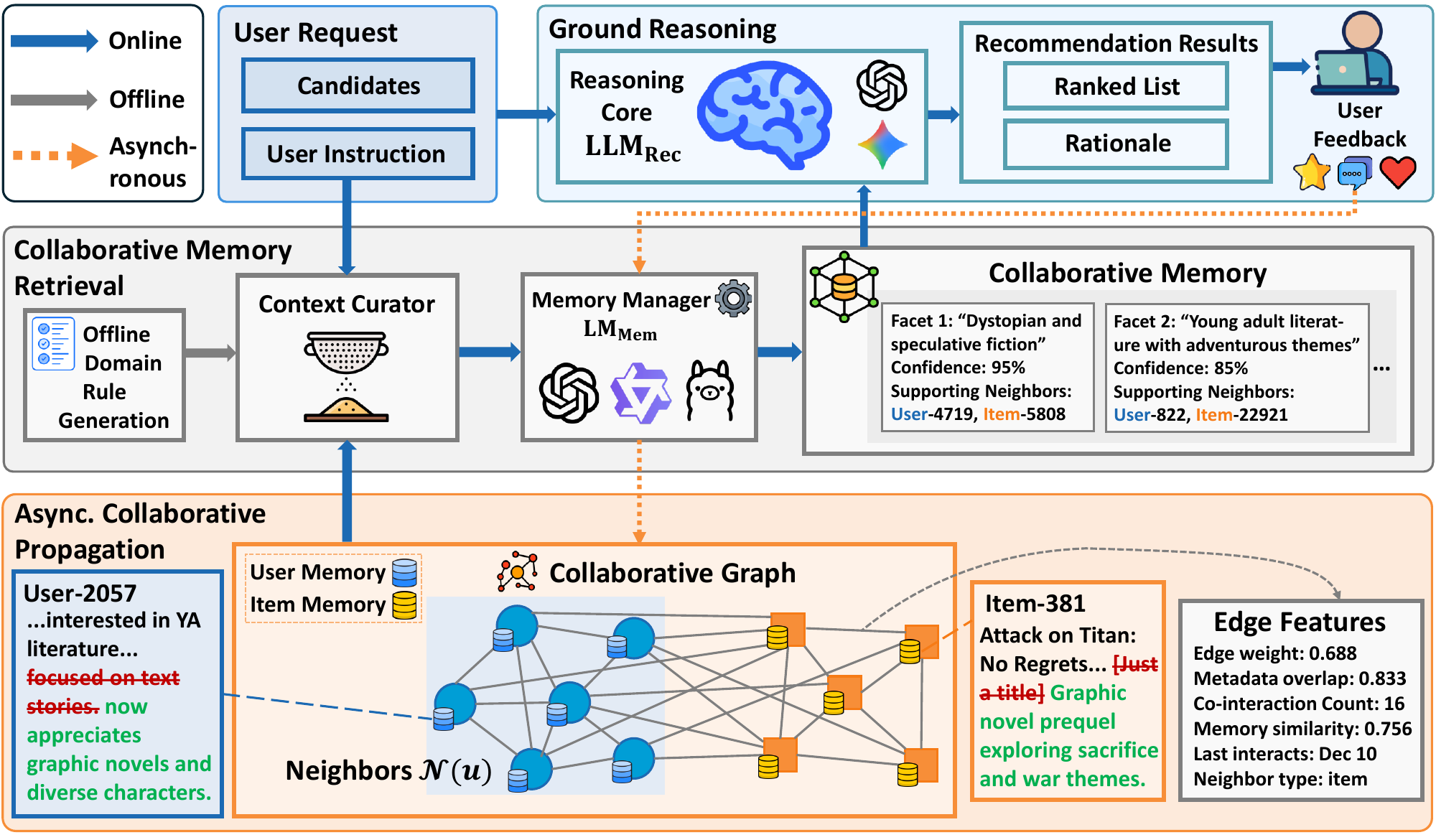} 
    \caption{The overall framework of \textbf{\texttt{MemRec}}, decoupling reasoning ($\text{LLM}_{\text{Rec}}$) from memory management ($\text{LM}_{\text{Mem}}$). The three-stage pipeline consists of: \textit{Collaborative Memory Retrieval}, synthesizing high-order connectivity context from memory graph; \textit{Grounded Reasoning}, scoring items based on instruction and context; and \textit{Asynchronous Collaborative Propagation}, evolving the semantic memory graph in the background.}
    \vspace{-3mm}
    \label{fig:memrec_model}

\end{figure*}

\section{Methodology}
\label{sec:methodology}
\vspace{-1mm}
\paragraph{Problem Formulation} Let $\mathcal{U}$ and $\mathcal{I}$ denote the sets of users and items, respectively. For each user $u \in \mathcal{U}$, we denote their historical interactions as $H_u$. Given a target user $u$, a natural language instruction $\mathcal{I}_u$ requiring semantic interpretation (e.g., specific constraints, complex goals), and a set of candidate items $C \subseteq \mathcal{I}$, the objective is to generate a ranked list of recommendations accompanied by grounded justifications.
\vspace{-1mm}
\paragraph{Memory in Agentic RS}
In agentic RS, memory serves as the persistent state, storing information in semantic form to evolve user understanding over time. While traditional systems maintain simple textual metadata, recent agentic architectures formalize these as individual semantic profiles, conceptualized as memory modules ($M_u$ or $M_i$)~\cite{xu2025iagent, zhang2024agentcf}. In practice, these memories manifest as evolving textual narratives summarizing a user's preferences or an item's characteristics based on historical contexts. During recommendation, a reasoning agent $\text{LLM}_{\text{Rec}}$ leverages these memories to perform the task.

Despite these advancements, existing agentic RS predominantly adhere to an isolated memory paradigm. They treat the collective memory $M$ merely as a disconnected set of individual narratives $\{M_u\} \cup \{M_i\}$. For instance, reasoning for user $u$ relies solely on their personal siloed memory $M_u$. This isolation excludes critical collaborative signals from the broader community, hindering the system's ability to leverage collective intelligence.


\subsection{The MemRec Pipeline}

To address this limitation, MemRec introduces a collaborative framework featuring an architecturally decoupled Memory Manager ($\text{LM}_{\text{Mem}}$). This manager operates on a unified memory graph $G = (\mathcal{V}, E)$. The node set $\mathcal{V} = \mathcal{U} \cup \mathcal{I}$ represents users and items, where each node $v \in \mathcal{V}$ stores its corresponding evolving semantic memory $M_v$. The edges $E$ encode interactions and derived relations connecting these memories. Unlike approaches relying solely on isolated node memories, MemRec leverages the high-order connectivity of $G$ to synthesize and propagate collaborative signals.

As illustrated in Figure \ref{fig:memrec_model}, MemRec operates in three key stages. Firstly, Collaborative Memory Retrieval processes the expansive graph to extract and synthesize a concise \textbf{Collaborative Memory} ($M_{\text{collab}}$) for the current task. Secondly, Grounded Reasoning utilizes this synthesized context to perform recommendations with enhanced grounding. Finally, Asynchronous Collaborative Propagation dynamically updates the individual semantic memories ($M_v$) across the graph, capturing emerging trends and shifting user preferences without disrupting the ongoing agentic interactions.

\subsubsection{Collaborative Memory Retrieval}
A central challenge in harnessing collaborative memory lies in mitigating \textit{cognitive overload} for the reasoning agent. Naively retrieving raw memories from all neighbors not only exceeds the context window constraints, but more crucially, bombards the LLM with noise, resulting in hallucinations and diminished instruction adherence. Our objective, therefore, is to extract a collaborative memory ($M_{\text{collab}}$) from the raw graph that maximizes relevance to the user's recommendation needs while rigorously filtering out extraneous interactions.

To this end, we draw conceptual inspiration from Information Bottleneck (IB) theory: our goal is to derive a compressed representation of the raw graph context that preserves maximal information relevant to the target task, while minimizing irrelevant or redundant signals. Guided by this insight, we adopt a "Curate-then-Synthesize" strategy. We first curate the raw collaborative graph by pruning redundant neighbors to reduce its complexity and size. Subsequently, we synthesize the distilled graph to amplify informative collaborative signals for the downstream reasoning agent. We elaborate on these two stages below.

\vspace{-2mm}

\paragraph{LLM-Guided Context Curation}
Conventional graph pruning strategies generally fall into two categories: (i) traditional rule-based heuristics, such as random walk-based methods~\cite{perozzi2014deepwalk}, which rely on predefined structural assumptions and lack semantic awareness; and (ii) fully learned neural scorers, such as GNN-based attention weights~\cite{velickovic2018graph}, which require expensive training and often lack interpretability. Both approaches present limitations for LLM-based agents. Heuristic methods cannot adapt to domain-specific semantic nuances, while learned scorers introduce significant computational overhead and integration complexity. 

To overcome these challenges, we propose a novel zero-shot \textbf{LLM-as-Rule-Generator} paradigm, which harnesses the rich background knowledge and semantic understanding of advanced LLMs to autonomously generate domain-specific curation rules. These rules are employed to guide the curation process, enabling efficient and adaptive collaborative memory construction tailored to the downstream LLM’s needs. Specifically, in an offline phase, $\text{LM}_{\text{Mem}}$ analyzes domain statistics $\mathcal{D}_{\text{domain}}$ (e.g., interaction density, category distribution) to synthesize interpretable heuristics.
\begin{equation}
\label{eq:rule_gen}
R_{\text{domain}} \leftarrow \text{LM}_{\text{Mem}}(\mathcal{D}_{\text{domain}} \| P_{\text{meta}}) \quad \text{(Offline)}
\end{equation}
Here, $P_{\text{meta}}$ acts as a generic meta-prompt guiding $\text{LM}_{\text{Mem}}$ to generate a set of domain-specific heuristic rules $R_{\text{domain}}$ tailored to balance relevance and diversity for the target dataset (see Appendix \ref{sec:prompts} and \ref{sec:generated-rules} for templates and examples). At inference time, these rules act as a high-speed filter, selecting the top-$k$ neighbors $N'_k(u)$ in milliseconds:
\begin{equation}
\label{eq:curation}
N'_k(u) = \text{Curate}(N(u), R_{\text{domain}}, k)
\end{equation}

This step acts as the first coarse ``compression'' pass in the IB framework, efficiently discarding neighbors with low potential mutual information. To illustrate the domain-adaptivity of this zero-shot generation, Figure \ref{fig:prompt_curation} presents a snippet of the generated rules for the Books dataset.

\paragraph{Collaborative Memory Synthesis}
The goal of this stage is to distill the raw information from curated neighbors $N'_k$ into a concise, structured format ($M_{\text{collab}}$) that maximizes informative signals for the downstream reasoning agent.
$\text{LM}_{\text{Mem}}$ synthesizes these signals into a set of structured preference facets $\{F\}$ as collaborative memory $M_{\text{collab}}$:
\begin{equation}
\label{eq:synthesis}
M_{\text{collab}} = \{F\} \leftarrow \text{LM}_{\text{Mem}}(\text{Rep}(N'_k) \| M_u^{t-1} \| P_{\text{synth}})
\end{equation}
where $\text{Rep}(N'_k)$ denotes the representation of neighbor information. To effectively synthesize signals within the LLM's limited context window, we adopt a tiered representation strategy. The target user $u$ is represented by their full, accumulated semantic memory $M_u^{t-1}$ to provide comprehensive background context.
Neighboring nodes in $N'_k$ are provided via compact contextual representations (e.g., condensed signals derived from memory or recent behaviors) designed to offer immediate evidence of collaborative patterns without overwhelming the model with verbose histories.
The synthesis prompt $P_{\text{synth}}$ (a snippet of which is illustrated in Figure \ref{fig:prompt_synthesis_snippet}, with the complete template available in Appendix \ref{sec:prompt_stage_r_synthesis}) then guides $\text{LM}_{\text{Mem}}$ to extract high-level facets from these tiered inputs. To make this concept concrete, Figure \ref{fig:collab_memory_example} contrasts a standard isolated memory with our synthesized collaborative memory facets, illustrating how \texttt{MemRec} enriches the reasoning context with structured, community-driven evidence.

\begin{figure}[t]
\centering
\begin{tcolorbox}[colback=gray!4, colframe=black, arc=0pt, boxrule=0.5pt, fontupper=\small\sffamily, left=2mm, right=2mm, top=1mm, bottom=1mm]
\textbf{LLM-Generated Curation Rules (Books Snippet)} \\[-1mm]
\rule{\linewidth}{0.4pt} \\[1mm]
\textbf{Rule 1: Content Similarity Boost} \\
$\bullet$ If \texttt{metadata\_overlap} > 0.6: Apply 2.5x multiplier. \\
$\bullet$ \textit{Rationale: Books are highly content-driven...} \\[1mm]
\textbf{Rule 2: Collaborative Filtering with Threshold} \\
$\bullet$ If \texttt{co\_interaction} > 3: Apply 1.8x multiplier. \\
$\bullet$ \textit{Rationale: Meaningful CF signal requires sufficient overlap.}
\end{tcolorbox}
\vspace{-2mm}
\caption{Snippet of the domain-adaptive curation rules generated by the $\text{LM}_{\text{Mem}}$ on the Books dataset.}
\label{fig:prompt_curation}
\vspace{-1mm}
\end{figure}

\begin{figure}[t]
\centering
\begin{tcolorbox}[colback=gray!4, colframe=black, arc=0pt, boxrule=0.5pt, fontupper=\small\sffamily, left=2mm, right=2mm, top=1mm, bottom=1mm]
\textbf{Prompt Snippet: Collaborative Synthesis} \\[-1mm]
\rule{\linewidth}{0.4pt} \\[1mm]
\textbf{Your Task:} Analyze the target user's personal memory and the raw collaborative memories from the curated neighboring users and items. Identify distinct preference facets that characterize this user's current interests. \\[1mm]
\textbf{Output Requirement:} For each facet, provide:\\
1. A concise description of the preference theme. \\
2. A confidence score (0-1). \\
3. A list of supporting neighbors providing evidence.
\end{tcolorbox}
\vspace{-2mm}
\caption{Snippet of the synthesis prompt used by $\text{LM}_{\text{Mem}}$ to distill collaborative memory facets.}
\label{fig:prompt_synthesis_snippet}
\vspace{-1mm}
\end{figure}

\begin{figure}[t]
\centering
\begin{tcolorbox}[colback=blue!4, colframe=black, arc=0pt, boxrule=0.5pt, fontupper=\small\sffamily, left=2mm, right=2mm, top=1mm, bottom=1mm]
\textbf{Isolated vs. Synthesized Collaborative Memory} \\[-1mm]
\rule{\linewidth}{0.4pt} \\[1mm]
\textcolor{gray}{\textbf{[Isolated $M_u^{t-1}$]:}} \textcolor{gray}{"User prefers dystopian settings. Recently interacted with 'The Hunger Games'."} \\[1mm]
\textcolor{blue!80!black}{\textbf{[Synthesized $M_{\text{collab}}$]:}} \\
$\bullet$ \textbf{Theme:} Cyberpunk \& Corporate Dystopia (Conf: 0.9) \\
\textit{Evidence:} User Neighbor (ID: 2057) shows deep interest in corporate control; Item Neighbor ('1984') shares foundational dystopian themes. \\
$\bullet$ \textbf{Theme:} High-Stakes Survival (Conf: 0.75) \\
\textit{Evidence:} Item Neighbor ('Battle Royale') exhibits strong survival elements matching recent user interactions.
\end{tcolorbox}
\vspace{-2mm}
\caption{Illustrative comparison of isolated vs. synthesized collaborative memory. The synthesized $M_{\text{collab}}$ extracts distinct themes with confidence scores and grounds them in specific neighbor evidence.}
\label{fig:collab_memory_example}
\vspace{-1mm}
\end{figure}

\subsubsection{Grounded Reasoning}
This stage is for reading the memory and performing the final ranking. By feeding the synthesized collaborative memory $M_{\text{collab}}$ alongside the user instruction $\mathcal{I}_u$ and the candidate item memories $C_{\text{info}}$, the $\text{LLM}_{\text{Rec}}$ executes the reasoning process:
\begin{equation}
\label{eq:rerank}
\{s_i, r_i\}_{i=1}^N \leftarrow \text{LLM}_{\text{Rec}}(\mathcal{I}_u \| M_{\text{collab}} \| C_{\text{info}} \| P_{\text{rerank}})
\end{equation}
The ranking prompt $P_{\text{rerank}}$ (Appendix \ref{sec:prompt_rerank}) instructs the LLM to generate a relevance score $s_i$ and a natural language rationale $r_i$ for each candidate based on the provided context. Grounding the reasoning in $M_{\text{collab}}$ ensures that the generated rationale is factually supported by broader community evidence provided.

\subsubsection{Async. Collaborative Propagation}
A static graph is inherently limited in its ability to capture evolving trends and shifting user preferences. As users continue to interact with items, the semantic representations stored within their corresponding memory nodes must be dynamically updated to reflect the most current patterns and preferences. Failure to adapt these representations risks diminishing the relevance and effectiveness of the recommender system over time. Drawing inspiration from Label Propagation algorithms \cite{zhu2002learning}, which spread information to connected nodes based on proximity within the graph structure, we introduce a mechanism to propagate "semantic labels" (insights) derived from new interactions asynchronously.

The update process conceptually involves two steps including updating the directly interacting nodes and propagating insights to neighbors. When user $u$ interacts with item $i_c$ at time step $t$, $\text{LM}_{\text{Mem}}$ first generates updates for the user's own memory $M_u^t$ and the item's memory $M_{i_c}^t$:
\begin{equation}
\label{eq:write_u}
M_u^{t}, M_{i_c}^{t} \leftarrow \text{LM}_{\text{Mem}}(M_{\text{collab}} \| M_u^{t-1} \| M_{i_c}^{t-1} \| P_{\text{update}})
\end{equation}
Here $M_u^{t}$ and $M_u^{t-1}$ denote the user's memory state at the current time step $t$ and the previous time step $t-1$, respectively. Crucially, this process also facilitates collaborative propagation by identifying connected neighbors from $N'_k(u)$ and propagating the shared theme as incremental updates $\Delta M_{\text{neigh}}$:
\begin{equation}
\label{eq:propagate}
\begin{split}
\{\Delta M_{\text{neigh}}\} \leftarrow \text{LM}_{\text{Mem}}(& M_{\text{collab}} \| M_u^{t-1} \| M_{i_c}^{t-1} \| \\
& N'_k(u) \| P_{\text{update}})
\end{split}
\raisetag{1.2\baselineskip}
\end{equation}
This explicit propagation enriches the global memory graph with high-order signals. To help visualize this evolution from an old memory state to a newly updated collaborative graph, Appendix \ref{sec:appendix_case_study} details a complete qualitative journey of this update process.

Crucially, we optimize this memory evolution to resolve the update bottleneck. While a naive synchronous approach scales linearly ($O(|N'_k|)$ calls) and incurs massive input token redundancy by repeating the user context for each neighbor, \texttt{MemRec} achieves an $O(1)$ Call Complexity per interaction. This is accomplished by decoupling the update frequency from the primary reasoning loop and batching reflections asynchronously. Specifically, we execute the logical steps of self-reflection (Eq. \ref{eq:write_u}) and neighbor propagation (Eq. \ref{eq:propagate}) as a single, batched asynchronous operation. A unified prompt $P_{\text{update}}$ (Appendix \ref{sec:prompt_stage_w}) guides the $\text{LM}_{\text{Mem}}$ to jointly synthesize all updates, ensuring continuous graph evolution without disrupting the online recommendation latency.

%% file: experiments.tex
\section{Empirical Evaluation}
\label{sec:experiments}

In this paper, we conduct extensive experiments to answer the following research questions:
\begin{itemize}[leftmargin=*, itemsep=1pt, topsep=1pt]

    \item \textbf{RQ1 (Overall Performance):} Does \texttt{MemRec} outperform SOTA baselines across diverse datasets?
    
    \item \textbf{RQ2 (Architectural Impact):} Is architectural decoupling crucial for information bottleneck?
    
    \item \textbf{RQ3 (Flexibility):} How cost-effective and flexible is \texttt{MemRec} for diverse deployments?
    
    \item \textbf{RQ4 (Ablation Study):} Are the core mechanisms of \texttt{MemRec} essential for its performance?

    \item \textbf{RQ5 (Robustness \& Quality):} How robust is \texttt{MemRec}, and what is its qualitative impact?
\end{itemize}
\vspace{-2mm}
\subsection{Experimental Setup}

\paragraph{Datasets}
We evaluate our methods on four widely used benchmark datasets covering diverse domains with varying interaction densities: \textbf{Amazon Books}, \textbf{Amazon Goodreads}, \textbf{MovieTV}, and \textbf{Yelp}.
For all datasets, we use the specific user instructions and evaluation splits provided by InstructRec \cite{xu2025iagent} to ensure fair comparison with instruction-following baselines.
Table \ref{tab:dataset_stats} summarizes the basic statistics of these datasets. Detailed descriptions of each dataset and its domain characteristics are provided in Appendix \ref{app:dataset_details}.

\begin{table}[h!]
\centering
\caption{Statistics of the datasets used in experiments.}
\label{tab:dataset_stats}
\resizebox{\columnwidth}{!}{%
\begin{tabular}{l|rrrrr}
\toprule
\textbf{Dataset} & $\mathbf{|U|}$ & $\mathbf{|I|}$ & $\mathbf{|E|}$ & $\mathbf{\bar{L}_u}$ & \textbf{Density} \\
\midrule
Books & 7.4K & 120.9K & 207.8K & 28.2 & 2.33e-4 \\
GoodReads & 11.7K & 57.4K & 618.3K & 52.7 & 9.19e-4 \\
MovieTV & 5.6K & 29.0K & 79.7K & 14.1 & 4.87e-4 \\
Yelp & 3.0K & 31.6K & 63.1K & 21.4 & 6.77e-4 \\
\bottomrule
\end{tabular}%
}
\end{table}
\vspace{-2mm}

\begin{table*}[t]
\centering
\caption{Main results for \textbf{Books} and \textbf{Goodreads}. ``Improv.'' denotes the relative improvement of MemRec over the best baseline, and all improvements are statistically significant ($p < 0.05$).}
\label{tab:main_results_1}
\resizebox{\textwidth}{!}{%
\begin{tabular}{lrrrrrrrrrr} 
\toprule
& \multicolumn{5}{c}{\textbf{Books}} & \multicolumn{5}{c}{\textbf{Goodreads}} \\
\cmidrule(lr){2-6} \cmidrule(lr){7-11} 
\textbf{Model} & \textbf{H@1} & \textbf{H@3} & \textbf{N@3} & \textbf{H@5} & \textbf{N@5} & \textbf{H@1} & \textbf{H@3} & \textbf{N@3} & \textbf{H@5} & \textbf{N@5} \\
\midrule
LightGCN & 0.1753 & 0.3259 & 0.2596 & 0.5703 & 0.3592 & 0.2499 & 0.5879 & 0.4432 & \underline{0.7903} & 0.5263 \\
SASRec & 0.0914 & 0.2830 & 0.2001 & 0.4845 & 0.2824 & 0.1324 & 0.3518 & 0.2576 & 0.5407 & 0.3349 \\
P5 & 0.2192 & 0.3607 & 0.2994 & 0.5273 & 0.3671 & 0.1569 & 0.3229 & 0.2509 & 0.5060 & 0.3256 \\
\midrule
Vanilla LLM & 0.3138 & 0.5617 & 0.4533 & 0.7270 & 0.5226 & 0.2864 & 0.4662 & 0.3948 & 0.7390 & 0.5041 \\
iAgent & 0.3925 & 0.5560 & 0.4858 & 0.6905 & 0.5409 & 0.2617 & 0.4949 & 0.3954 & 0.6591 & 0.4626 \\
\midrule
RecBot & 0.3984 & 0.5491 & 0.4846 & 0.6786 & 0.5376 & 0.2705 & 0.4754 & 0.3876 & 0.6495 & 0.4589 \\
AgentCF & 0.3457 & 0.6060 & 0.4960 & 0.7403 & 0.5512 & 0.2951 & 0.5910 & 0.4654 & 0.7726 & 0.5399 \\
i$^2$Agent & \underline{0.4453} & \underline{0.6517} & \underline{0.5649} & \underline{0.7708} & \underline{0.6138} & \underline{0.3099} & \underline{0.6079} & \underline{0.4825} & 0.7675 & \underline{0.5481} \\
\midrule
\rowcolor{gray!20}
MemRec & \textbf{0.5117} & \textbf{0.6915} & \textbf{0.6152} & \textbf{0.8007} & \textbf{0.6601} & \textbf{0.3997} & \textbf{0.6658} & \textbf{0.5540} & \textbf{0.8052} & \textbf{0.6112} \\
\textit{Improv.} & \textit{+14.91\%} & \textit{+6.11\%} & \textit{+8.90\%} & \textit{+3.88\%} & \textit{+7.54\%} & \textit{+28.98\%} & \textit{+9.52\%} & \textit{+14.82\%} & \textit{+1.89\%} & \textit{+11.51\%} \\
\bottomrule
\end{tabular}%
}
\end{table*}

\begin{table*}[t]
\centering
\caption{Main results for \textbf{MovieTV} and \textbf{Yelp}. Notation follows Table~\ref{tab:main_results_1}; all improvements are significant ($p < 0.05$).}
\label{tab:main_results_2}
\resizebox{\textwidth}{!}{%
\begin{tabular}{lrrrrrrrrrr} 
\toprule
& \multicolumn{5}{c}{\textbf{MovieTV}} & \multicolumn{5}{c}{\textbf{Yelp}} \\
\cmidrule(lr){2-6} \cmidrule(lr){7-11} 
\textbf{Model} & \textbf{H@1} & \textbf{H@3} & \textbf{N@3} & \textbf{H@5} & \textbf{N@5} & \textbf{H@1} & \textbf{H@3} & \textbf{N@3} & \textbf{H@5} & \textbf{N@5} \\
\midrule
LightGCN & 0.3482 & 0.5643 & 0.4738 & 0.6883 & 0.5241 & 0.3444 & 0.5658 & 0.4720 & 0.7546 & 0.5494 \\
SASRec & 0.3399 & 0.5233 & 0.4470 & 0.6382 & 0.4942 & 0.2305 & 0.4312 & 0.3458 & 0.5597 & 0.3980 \\
P5 & 0.1696 & 0.3206 & 0.2554 & 0.5008 & 0.3290 & 0.1444 & 0.3207 & 0.2435 & 0.5220 & 0.4785 \\
\midrule
Vanilla LLM & 0.4050 & 0.7764 & 0.6098 & 0.8603 & 0.6445 & 0.1692 & 0.5275 & 0.3696 & 0.6861 & 0.4360 \\
iAgent & 0.4253 & 0.6170 & 0.5361 & 0.7420 & 0.5871 & 0.3995 & 0.6005 & 0.5148 & 0.7300 & 0.5681 \\
\midrule
RecBot & 0.4367 & 0.6113 & 0.5375 & 0.7309 & 0.5866 & 0.4007 & 0.6003 & 0.5156 & 0.7169 & 0.5636 \\
AgentCF & 0.3906 & 0.6693 & 0.5523 & 0.7864 & 0.6006 & 0.1925 & 0.4374 & 0.3326 & 0.6374 & 0.4147 \\
i$^2$Agent & \underline{0.4912} & \underline{0.7225} & \underline{0.6262} & \underline{0.8221} & \underline{0.6672} & \underline{0.4205} & \underline{0.6454} & \underline{0.5517} & \underline{0.7648} & \underline{0.6007} \\
\midrule
\rowcolor{gray!20}
MemRec & \textbf{0.5882} & \textbf{0.7819} & \textbf{0.7011} & \textbf{0.8817} & \textbf{0.7422} & \textbf{0.4868} & \textbf{0.6912} & \textbf{0.6053} & \textbf{0.7908} & \textbf{0.6463} \\
\textit{Improv.} & \textit{+19.75\%} & \textit{+8.22\%} & \textit{+11.96\%} & \textit{+7.25\%} & \textit{+11.24\%} & \textit{+15.77\%} & \textit{+7.10\%} & \textit{+9.72\%} & \textit{+3.40\%} & \textit{+7.59\%} \\
\bottomrule
\end{tabular}%
}
\end{table*}

\paragraph{Baselines}
We evaluate MemRec against a suite of strong baselines, grouped by their underlying memory paradigms. The first category comprises traditional pre-LLM methods that utilize dense latent embeddings to encode and preserve historical information, including LightGCN~\cite{he2020lightgcn}, SASRec~\cite{kang2018self}, and P5~\cite{geng2022recommendation}. The second category encompasses memory-based approaches developed in the era following AgentRS, which can be further subdivided into: (1) models with \textit{no explicit memory}, such as Vanilla LLM \cite{liu2023chatgpt} that operate on raw interaction histories; (2) those employing \textit{static memory}, exemplified by iAgent’s fixed profile representations; and (3) \textit{dynamic memory} agents that update isolated memories, namely i$^2$Agent \cite{xu2025iagent}, AgentCF \cite{zhang2024agentcf}, and RecBot \cite{tang2025interactive}. In contrast, our \textbf{\texttt{MemRec}} introduces a new paradigm, \textbf{Dynamic Collaborative Memory}, featuring asynchronous graph propagation. Baseline details are in Appendix \ref{app:baseline_details}.
\vspace{-2mm}
\paragraph{Experimental Setup}
\label{sec:experimental_setup}
We implement \texttt{MemRec} using \textbf{gpt-4o-mini}~\cite{openai2024hello} for both $\text{LLM}_{\text{Rec}}$ and $\text{LM}_{\text{Mem}}$, setting $k=16$ and $N_f=7$.
For main results, we set the candidate list size $N=10$ on full test sets and observe consistent trends with larger candidate sets in Appendix \ref{sec:appendix_k20}. We report \textbf{Hit Rate (H@K)} and \textbf{NDCG (N@K)} for $K \in \{1, 3, 5\}$.
Following~\cite{zhang2024agentcf}, we utilize a randomly sampled subset of 1000 users in subsequent studies. More comprehensive implementation details are provided in Appendix \ref{sec:implementation_details}.

\subsection{Main Results (RQ1)}

Tables \ref{tab:main_results_1} and \ref{tab:main_results_2} present the comprehensive performance comparison across four datasets. All reported improvements of MemRec over the best baseline are statistically significant ($p < 0.05$). From the results, we observe some key findings:

\begin{itemize}[leftmargin=*, itemsep=1pt, topsep=1pt]
    \item 
    \textbf{\texttt{MemRec} decisively outperforms all baselines in ranking metrics.} 
    Our framework achieves state-of-the-art performance across all reported metrics on the four benchmark datasets. Notably, on Goodreads, \texttt{MemRec} achieves its most significant gain, improving H@1 by +28.98\% relative to the strongest baseline i$^2$Agent. On the dense Yelp dataset, it also demonstrates superiority across diverse metrics (e.g., +15.77\% in H@1 and +7.59\% in N@5), proving its ability to effectively capture broad community signals via collaborative memory combined with specific user instructions.

    \item 
    \textbf{Memory paradigm hierarchy: Collaborative > Dynamic > Static > No Memory.} 
    Among the agentic baselines, dynamic memory approaches consistently outperform static memory methods such as iAgent. In turn, static methods generally achieve better results than approaches with no explicit memory (Vanilla LLM). While these findings align with current trends, we observe that even SOTA dynamic agents (e.g., AgentCF) still significantly underperform relative to \texttt{MemRec}. This underscores the limitation of considering memories in isolation and highlights the absolute necessity of explicitly injecting collaborative signals into the agent's memory module.

    \item 
    \textbf{\texttt{MemRec} bridges the gap between traditional CF robustness and modern LLM reasoning.} 
    Traditional models like LightGCN show inconsistent performance, struggling on sparse tasks (Books) while remaining competitive on dense graphs (Yelp). Conversely, older LLM paradigms like P5 struggle due to limited model capacity and reliance on ID-based pre-training. \texttt{MemRec} successfully bridges these worlds, leveraging powerful LLM reasoning to dominate where traditional CF fails, while using collaborative graph signals to surpass isolated agentic baselines.
\end{itemize}

\begin{figure*}[t]
    \centering
    \includegraphics[width=\textwidth]{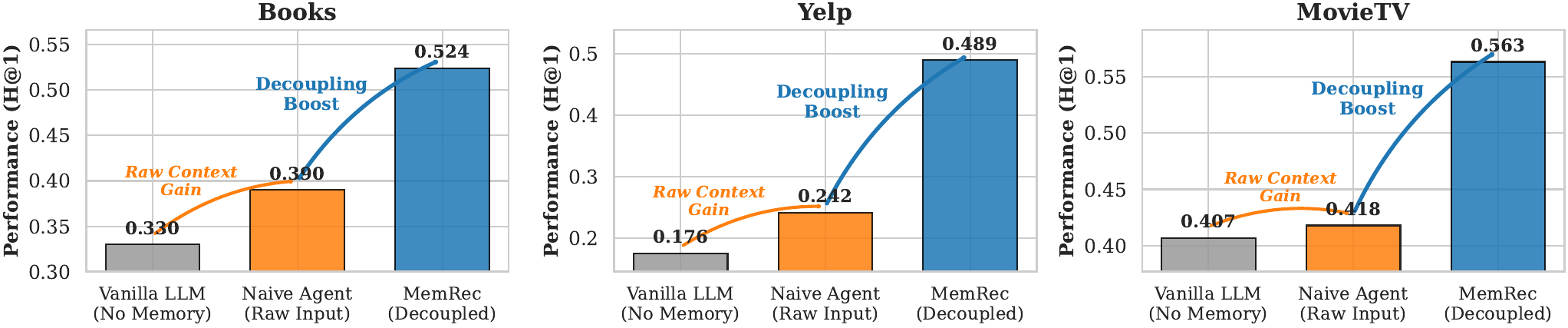}
    \caption{Impact of architectural decoupling on H@1. \textbf{MemRec} (\protect\textcolor{blue}{blue}) overcomes the information bottleneck that causes \textbf{Naive Agents} (\protect\textcolor{orange}{orange}) to plateau, achieving substantial gains over both Naive and \textbf{Vanilla LLM} (\protect\textcolor{gray}{gray}). 
    }
    \vspace{-1mm}
    \label{fig:architectural_impact}
\end{figure*}

\subsection{Impact of Cognitive Overload (RQ2)} \label{sec:architectural_analysis}
Cognitive overload occurs when agents fail to distill pertinent signals from raw graph contexts. We validate MemRec's architecture by comparing it against a Vanilla LLM and a Naive Collaborative Agent (which processes uncurated context in one stage) in Figure~\ref{fig:architectural_impact}. Results show the Naive Agent (orange) plateaus because a single model cannot effectively ingest verbose context and perform complex ranking simultaneously, creating an information bottleneck. MemRec (blue) breaks this plateau through architectural decoupling. By separating memory management ($\text{LM}_{\text{Mem}}$) from reasoning ($\text{LLM}_{\text{Rec}}$), it ensures the ranker receives only high-signal, "Curate-then-Synthesize" context. Consequently, MemRec consistently and substantially outperforms the monolithic approach across all datasets (e.g., +34\% relative H@1 gain on Books).

\vspace{-1mm}

\subsection{Flexibility and Cost-Effectiveness (RQ3)}
\label{sec:flexibility_analysis}

To evaluate the flexibility and practical deployment viability of \texttt{MemRec}, we analyze the trade-offs between reasoning performance, sequential latency, and computational cost across various configurations (Figure~\ref{fig:efficiency_bubble}). A detailed metrics breakdown is provided in Appendix~\ref{sec:app_efficiency_analysis}.


Crucially, this analysis focuses on the \textbf{online inference cost} (i.e., the Re-ranking stage). A fundamental advantage of \texttt{MemRec} is its ability to offload the computationally heavy cognitive load of processing dense collaborative graphs to asynchronous offline batches. Unlike monolithic baseline agents that must ingest vast raw contexts in real-time, \texttt{MemRec} achieves superior online efficiency by only querying the distilled collaborative memory. This architectural shift establishes a vastly superior Pareto frontier. Specifically, the \textbf{Standard} (\texttt{gpt-4o-mini}) and \textbf{Cloud-OSS} (a high-throughput open-weights model on Azure matching the mini price tier) configurations achieve near-ceiling performance at a fraction of the baselines' online cost, demonstrating model-agnostic serving efficiency. Furthermore, the \textbf{Vector} variant showcases extreme modularity by replacing the LLM ranker with lightweight vector similarity search to achieve sub-millisecond online latency. Additionally, as detailed in Appendix Table \ref{tab:architectural_analysis_app}, \textbf{Local} deployments (e.g., \texttt{Qwen-2.5-7B}) provide competitive performance for privacy-sensitive domains.
\vspace{-2mm}
\begin{figure}[t]
    \centering
    \includegraphics[width=\columnwidth]{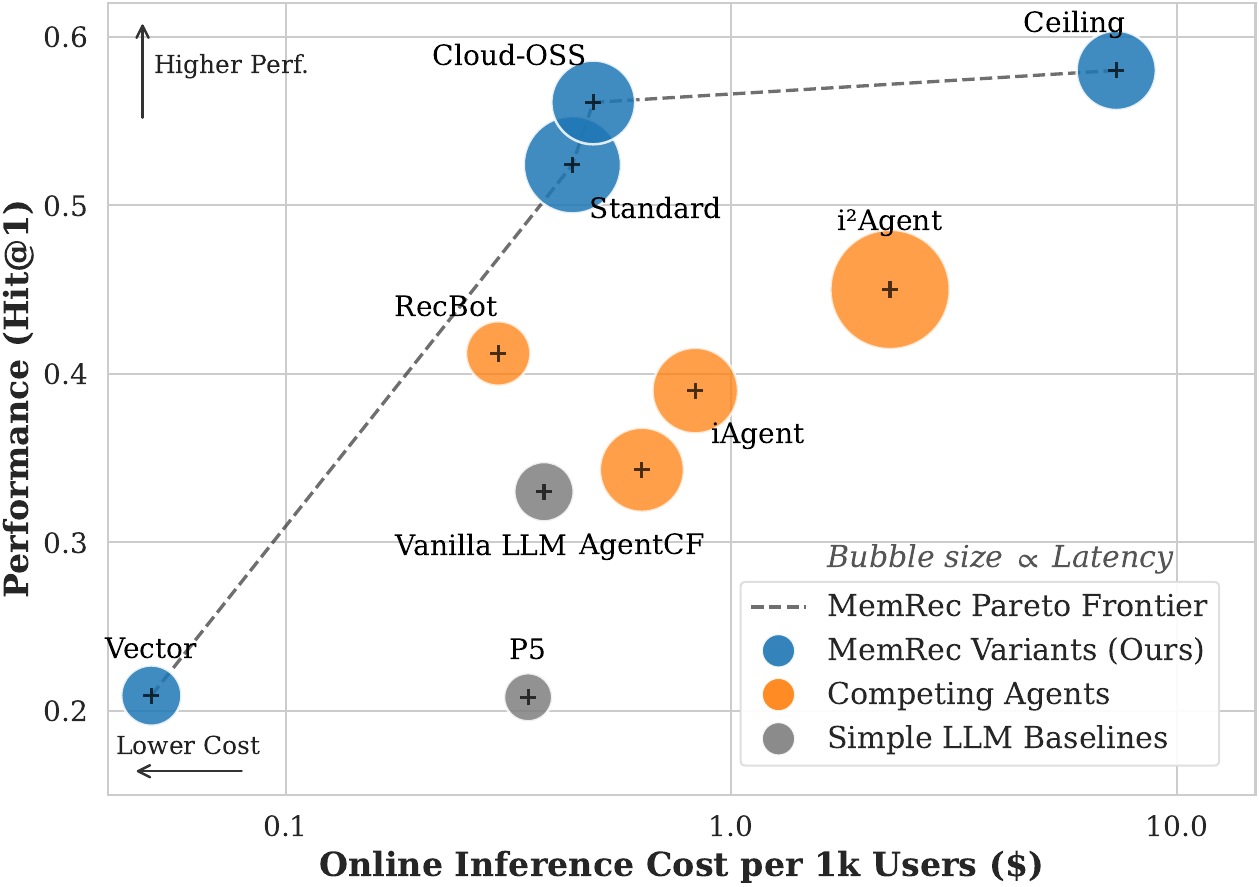}
    \caption{Efficiency-Cost-Performance Landscape. 
    }
    \label{fig:efficiency_bubble}
    \vspace{-1mm}
\end{figure}

\subsection{Ablation Studies (RQ4)}
\label{sec:ablation}

A comprehensive ablation study (Table \ref{tab:ablation_all}) confirms the positive contribution of MemRec's key collaborative components. Removing the collaborative retrieval stage (\textit{w/o Collab. Read}), where the agent only reflects on isolated personal history, causes a drastic 9.9\% drop in H@1, validating the critical role of synthesizing global graph signals over relying solely on isolated memory. Replacing the domain-adaptive LLM curator with generic heuristic rules (\textit{w/o LLM Curation}) leads to a 5.5\% drop, confirming the superior precision of our zero-shot, LLM-guided curation strategy in filtering noise. Finally, disabling asynchronous propagation (\textit{w/o Collab. Write}) results in a 4.2\% drop. This suggests that while a static graph supports broad retrieval (high H@5), dynamic collaborative updates are crucial for refining top-tier ranking precision (H@1) by capturing evolving community trends.

\begin{table}[h!]
\centering
\caption{Comprehensive ablation study on \texttt{books}.
}
\vspace{-2mm}
\label{tab:ablation_all}
\resizebox{\columnwidth}{!}{%
\begin{tabular}{l|ccccc|r}
\toprule
\textbf{Model Config.} & \textbf{H@1} & \textbf{H@3} & \textbf{N@3} & \textbf{H@5} & \textbf{N@5} & \textbf{Drop} \\
\midrule
\rowcolor{gray!10} \textbf{MemRec (Full)} & \textbf{0.527} & \textbf{0.713} & \textbf{0.634} & 0.803 & \textbf{0.670} & - \\
\midrule
w/o Collab. Write & 0.505 & 0.702 & 0.619 & \textbf{0.814} & 0.665 & 4.2\% \\
w/o LLM Curation & 0.498 & 0.685 & 0.606 & 0.788 & 0.648 & 5.5\% \\
w/o Collab. Read & 0.475 & 0.650 & 0.575 & 0.769 & 0.624 & 9.9\% \\
\bottomrule
\end{tabular}%
}
\end{table}

\subsection{Further Analysis (RQ5)}
\label{sec:further_analysis}

\begin{figure}[h]
    \centering
    \includegraphics[width=1.00\columnwidth]{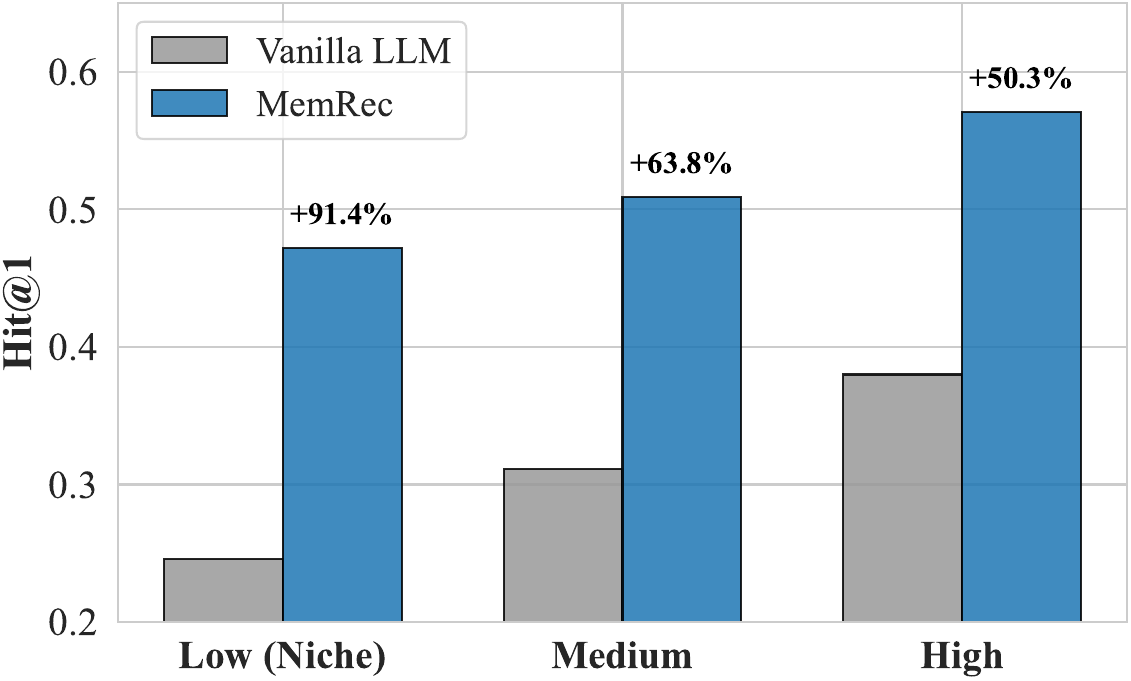}
    \caption{Sub-group performance across activity levels. \texttt{MemRec} yields massive gains (+91.4\%) for data-sparse niche users over the Vanilla baseline.}
    \label{fig:niche_user}
\end{figure}

\paragraph{Robustness to Highly Niche Users} 
Collaborative methods often struggle with "niche" users lacking sufficient interaction history. We stratified a 1,000-user test subset by activity level (Low, Medium, High). Rather than degrading, \texttt{MemRec} yields the greatest benefits for data-sparse users (Figure \ref{fig:niche_user}). By leveraging the collaborative graph to offset personal history deficits, \texttt{MemRec} achieves a massive \textbf{+91.4\%} relative Hit@1 improvement over the Vanilla LLM for the Low activity group.

\begin{figure}[h]
    \centering
    \includegraphics[width=0.95\columnwidth]{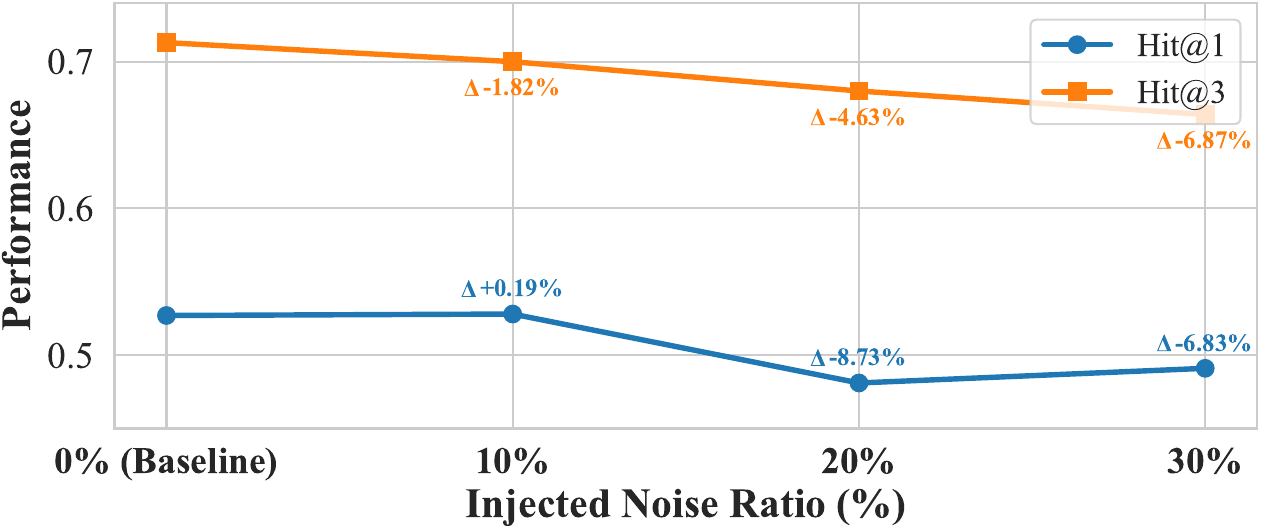}
    \caption{Perturbation analysis under varying noise injection ratios. \texttt{MemRec} exhibits minimal degradation, demonstrating strong resistance to noisy neighbors.}
    \label{fig:noise_robustness}
\end{figure}

\paragraph{Perturbation Analysis} 
To evaluate the risk of negative propagation, we injected random noise (fake items) into target user histories. As shown in Figure \ref{fig:noise_robustness}, \texttt{MemRec} remains highly robust, maintaining a competitive 0.491 Hit@1 even under 30\% noise injection. This resilience stems from the "Curate-then-Synthesize" pipeline: the zero-shot LLM curation acts as a robust buffer, successfully filtering out irrelevant peers before they overwhelm the downstream reasoning agent.

\begin{figure}[t]
    \centering
    \includegraphics[width=1.00\columnwidth]{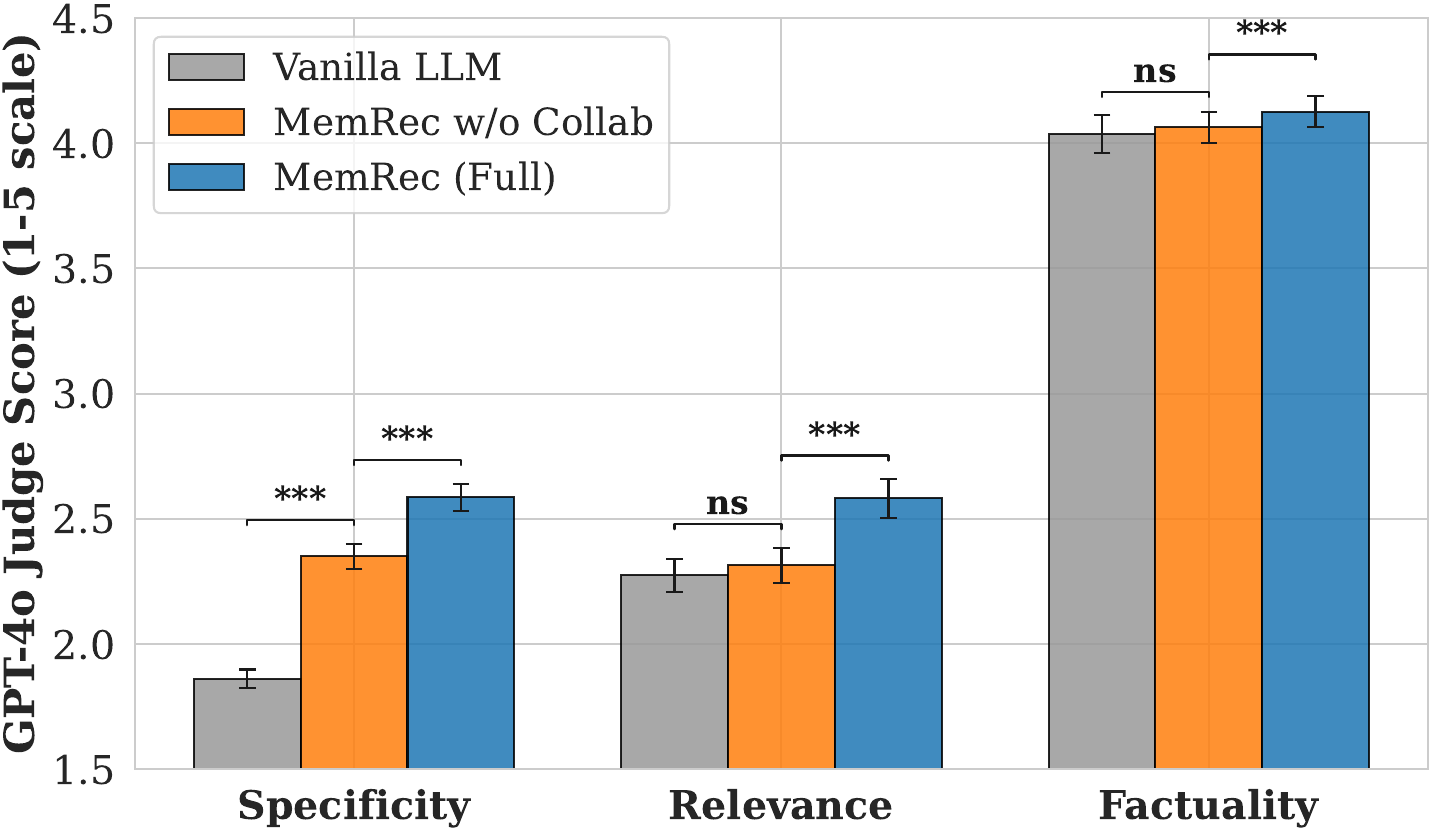}
    \caption{
    Rationale Quality Evaluation (GPT-4o Judge, 1-5 scale). Error bars show 95\% CIs; *** denotes $p<0.001$, while ns means not significant on paired t-test. 
    }
    \vspace{-2mm}
    \label{fig:rationale_eval}
\end{figure}

\paragraph{Rationale Quality Analysis}
A GPT-4o-based evaluation (Figure \ref{fig:rationale_eval}) shows MemRec significantly improves rationale \textit{Specificity} and \textit{Relevance} over baselines by incorporating collaborative signals. See Appendix \ref{sec:app_rationale_analysis} for full details.

\begin{figure}[h!]
    \centering
    \includegraphics[width=0.95\columnwidth]{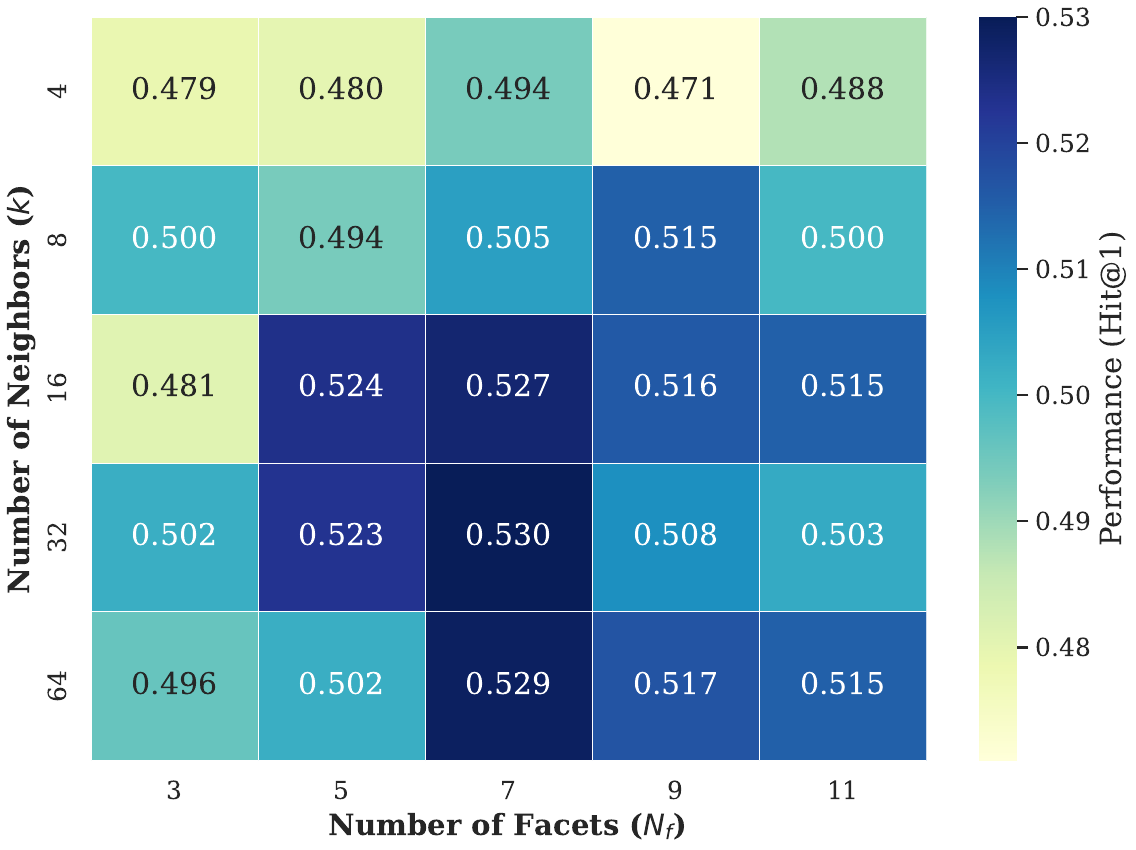}
    \caption{Hyperparameter sensitivity on \texttt{books}.
    }
    \label{fig:hyperparam_heatmap_h1}
    \vspace{-3mm}
\end{figure}

\paragraph{Hyperparameter Sensitivity}
We vary neighbors $k$ and facets $N_f$ in Figure \ref{fig:hyperparam_heatmap_h1} (Hit@1), observing a performance "sweet spot" around $k \in \{16, 32\}$ and $N_f=7$. Comprehensive analysis across full metrics is detailed in Appendix \ref{sec:appendix_hyper}.

\paragraph{Qualitative Analysis}
A comprehensive case study illustrating the complete collaborative journey including collaborative memory synthesis, grounded reasoning, and asynchronous memory propagation, is provided in Appendix \ref{sec:appendix_case_study}.

%% file: related_work.tex
\section{Related Works}
\label{sec:related_work}
\vspace{-1mm}
To overcome LLM context constraints for long-horizon tasks, research has evolved from basic Retrieval-Augmented Generation (RAG) pipelines \cite{lewis2020retrieval} to sophisticated, dedicated memory architectures. Systems like \textit{MemGPT} \cite{packer2023memgpt} and \textit{Generative Agents} \cite{park2023generative} demonstrate how decoupled memory managers and reflective synthesis can maintain long-term coherence. However, these general-purpose frameworks typically target factual or conversational domains, fundamentally neglecting the specialized, high-order connectivity required for collaborative recommendation environments.

In the realm of Agentic RS, approaches have transitioned from stateless prompting \cite{liu2023chatgpt} to incorporating explicit, dynamic memory. While recent state-of-the-art agents like \textit{i$^2$Agent} \cite{xu2025iagent} and simulation frameworks like \textit{AgentCF} \cite{zhang2024agentcf} employ self-reflection mechanisms to evolve user understanding over time, they remain paradigm-bound to isolated memory. Updates are strictly confined to the interacting user or item silos, failing to leverage the global collaborative signals that are vital for effective recommendation. MemRec addresses this critical gap by shifting the paradigm from isolated, self-reflective memory to a dynamic, collaborative memory graph. A more detailed review of related literature is provided in Appendix \ref{app:detailed_related_work}.

%% file: conclusion.tex
\vspace{-1mm}
\section{Conclusion}
\label{sec:conclusion}
Existing agent-based recommender systems typically rely on isolated user and item memories, which inherently restricts their ability to leverage broader community signals and accurately infer preferences, particularly for data-sparse users. To address this limitation, we identify \textit{collaborative memory} as a crucial design paradigm and practically instantiate it through \textbf{\texttt{MemRec}}. By architecturally decoupling high-level reasoning ($\text{LLM}_{\text{Rec}}$) from memory management ($\text{LM}_{\text{Mem}}$), \texttt{MemRec} successfully resolves the dual challenges of naïve collaborative approaches: it mitigates cognitive overload via zero-shot LLM-guided context curation and circumvents prohibitive update bottlenecks via asynchronous graph propagation. Extensive experiments demonstrate that this paradigm not only delivers substantial ranking performance gains across four diverse benchmarks, but also establishes a superior Pareto frontier that balances reasoning quality, online inference cost, and deployment flexibility. Future work will explore scaling \texttt{MemRec} to web-scale graphs and investigating privacy-preserving federated memory updates.

%% file: limitations.tex
\section{Limitations}
\label{sec:limitations}

Despite its strong performance and flexible architecture, \texttt{MemRec} has limitations that warrant future investigation. Currently, our asynchronous collaborative propagation is restricted to immediate neighbors to manage computational overhead; extending this to multi-hop community updates without introducing noise requires more efficient selection mechanisms. Furthermore, our context curation rules are derived from static domain statistics generated offline, which may need online adaptation to maintain efficacy in highly dynamic environments (e.g., news). Finally, while memory operations can be successfully offloaded to local models, achieving ceiling reasoning performance still relies on powerful proprietary LLMs, motivating future work on fully open-source stacks.

%% file: acknowledgements.tex
\section{Acknowledgements}
\label{sec:acknowledgements}

This work is supported by NSFC/RGC Joint Research Scheme (N\_HKBU214/24).

%% file: appendix.tex


% ...


\appendix





\section{Experimental Setup and Implementation Details}
\label{sec:appendix_implementation}

\subsection{Dataset Details}
\label{app:dataset_details}

We utilize four datasets widely used in recommendation research, encompassing diverse domains such as e-commerce, social reading, entertainment, and local services. As mentioned in Section \ref{sec:experiments}, we adopt the versions of these datasets augmented with natural language user instructions from \textbf{InstructRec} \cite{xu2025iagent}. The original data sources and their detailed descriptions are provided below:

\paragraph{Books}
Derived from the \textbf{Amazon review dataset}\footnote{\url{https://cseweb.ucsd.edu/~jmcauley/datasets/amazon_v2/}} \cite{ni2019justifying}, this subset focuses on book recommendations. It is characterized by incredibly sparse interactions and a vast item space. User preferences in this domain are typically stable and highly content-driven, focusing on specific genres, authors, or themes.

\paragraph{Goodreads}
Collected from the \textbf{Goodreads} social book cataloging website\footnote{\url{https://cseweb.ucsd.edu/~jmcauley/datasets/goodreads.html}} \cite{wan2019fine}, this dataset is notably dense compared to others. It features strong community interactions and rich metadata about books, including series information. Users on Goodreads often exhibit series-aware reading behaviors and are influenced by social signals.

\paragraph{MovieTV}
Also originating from the \textbf{Amazon review dataset} \cite{ni2019justifying}, this dataset covers movies and TV shows. The domain is marked by volatile user preferences often influenced by immediate context or trending content. While metadata like genre and cast are important, item recency frequently plays a critical role in user decision-making.

\paragraph{Yelp}
Sourced from the \textbf{Yelp Dataset}\footnote{\url{https://www.kaggle.com/datasets/yelp-dataset/yelp-dataset}}, this dataset consists of reviews for local businesses like restaurants and services. It is characterized by strong categorical constraints (e.g., cuisine type) and the critical importance of attributes like price range and location. User preferences here are often highly context-dependent.

\subsection{Baseline Model Details}
\label{app:baseline_details}

This appendix provides detailed descriptions of the baseline models used in our comparative evaluation. Following the categorization in the main text, we group these baselines based on their underlying memory paradigms into two major categories: traditional pre-LLM methods using latent embeddings, and memory-based approaches developed in the post-AgentRS era using semantic memory.

\subsubsection{Traditional Pre-LLM Methods (Latent Embeddings)}

These models represent the conventional paradigm where historical information is encoded and preserved using dense latent vectors, without explicit semantic memory structures for reasoning agents.

\begin{itemize}[leftmargin=*]
    \item \textbf{LightGCN} \cite{he2020lightgcn}: A state-of-the-art graph collaborative filtering model that simplifies the Graph Convolutional Network (GCN) design by removing feature transformation and nonlinear activation. It learns user and item embeddings by linearly propagating them on the user-item interaction graph, capturing high-order collaborative signals through structural connections.

    \item \textbf{SASRec} \cite{kang2018self}: A leading sequential recommendation model based on the self-attention mechanism. It models the entire user sequence to capture long-term semantics and dynamic dependencies, using an attention mechanism to selectively focus on relevant items in the history for making predictions.

    \item \textbf{P5} \cite{geng2022recommendation}: A unified framework that formulates various recommendation tasks as sequence-to-sequence language modeling problems. It utilizes a pre-trained T5 backbone and represents users and items as sequence tokens (IDs) within personalized prompts. While LLM-based, the original P5 relies on pre-trained knowledge related to these IDs and does not incorporate an evolving, descriptive memory component.
\end{itemize}

\subsubsection{Memory-based Approaches (Post-AgentRS Era)}

This category encompasses approaches developed in the era following AgentRS, utilizing LLMs with varying degrees of semantic memory capabilities.

\paragraph{(1) Models with No Explicit Memory}
These models operate by directly processing raw interaction histories without maintaining a persistent, structured semantic memory store.

\begin{itemize}[leftmargin=*]
    \item \textbf{Vanilla LLM (Zero-Shot Prompting)} \cite{liu2023chatgpt}: This baseline represents the direct application of a powerful instruction-tuned LLM (e.g., GPT-4o-mini) via API calls. For each prediction, the user's entire sequence of historical interactions is converted into a natural language string and fed into the LLM as a static context prompt. The model performs zero-shot selection from candidate items based solely on this provided raw history, serving as a baseline to measure the LLM's inherent capabilities independent of designed memory architectures.
\end{itemize}

\paragraph{(2) Static Memory Agents}
These agents utilize descriptive semantic information about users and items, but this "memory" remains fixed as a static context during inference and does not evolve.

\begin{itemize}[leftmargin=*]
    \item \textbf{iAgent} \cite{xu2025iagent}: An LLM-based autonomous agent designed for recommendation. It employs a static profile for each user, constructed from their historical interactions and available descriptive data. This fixed profile is fed into the LLM as context to generate recommendations. The key characteristic is that its understanding of the user does not adapt over time after initial construction.
\end{itemize}

\paragraph{(3) Dynamic Memory Agents (Isolated Updates)}
These agents possess a dynamic memory mechanism, allowing them to reflect on interactions and update their understanding. However, these updates are isolated to the individual agent and do not propagate collaboratively.

\begin{itemize}[leftmargin=*]
    \item \textbf{i$^2$Agent} \cite{xu2025iagent}: An extension of iAgent that introduces a "reflection" mechanism. After recommendations, the agent can reflect on user feedback to refine its internal state or strategy for future interactions. While dynamic, these reflections are confined to the individual agent's experience with a specific user.

    \item \textbf{AgentCF} \cite{zhang2024agentcf}: An agent-based collaborative filtering framework that simulates user-item interactions. Agents representing users and items can autonomously interact, learn from these interactions, and update their own preferences or characteristics. The memory update is dynamic but remains localized to the individual agents involved in the direct interaction.

    \item \textbf{RecBot} \cite{tang2025interactive}: A conversational recommender system that uses an LLM to engage with users. It maintains a dynamic dialogue history and can update its understanding of user preferences based on the ongoing conversation. This dynamic memory allows for multi-turn interactions but is limited to the context of the current user session.
\end{itemize}

\subsection{Implementation Details}
\label{sec:implementation_details}

\paragraph{Model Deployment}
We utilize a diverse set of models across different configurations. For proprietary models, we access \texttt{gpt-4o-mini} (Standard config.) and \texttt{gpt-4o} (Ceiling config.) via the Microsoft Azure OpenAI Service, using API version \texttt{2024-08-01-preview}. For the Cloud-OSS configuration, we employ the large-scale open-source \textbf{\texttt{gpt-oss-120b}} model via Azure Serverless APIs. For local open-source ablations (\texttt{Qwen-2.5-7B-Instruct}, \texttt{Meta-Llama-3-8B-Instruct}), we deploy them locally using the \textbf{vLLM} library~\cite{kwon2023efficient} for optimized high-throughput inference in \textbf{FP16 (half-precision)} mode. For the Vector configuration, we utilize the \texttt{all-MiniLM-L6-v2} Sentence Transformer~\cite{reimers2019sentence}. We used Gemini to polish sentences and improve language flow only. The core ideas, research, and results are fully our own work.

\paragraph{Hardware Environment}
All local experiments (specifically for Local-Qwen and Local-Llama) were conducted on a workstation equipped with a single \textbf{NVIDIA RTX A5000 GPU (24GB VRAM)}. While our local setup exhibits higher latency compared to optimized cloud APIs (as detailed in Table \ref{tab:architectural_analysis_app}), deploying 7B models on enterprise-grade inference hardware would likely yield competitive speeds.

\paragraph{Hyperparameters}
We set the neighbor count $k=16$ and the number of synthesis facets $N_f=7$. The max token budget for context retrieval is set to $\tau=1800$. We use a temperature of $0.0$ for all LLM calls to ensure reproducibility.

\paragraph{Neighbor Representation Strategy}
To efficiently manage the strict context token budget ($\tau=1800$) while maintaining broad neighbor coverage ($k=16$) during Stage-R, we implemented a practical tiered representation strategy. While item neighbors utilize their truncated semantic memory (initialized by metadata descriptions), user neighbors are represented by their sequence of recent interactions (e.g., titles of the last three acted items). This acts as a dense, token-efficient proxy for immediate interests, enabling the inclusion of diverse collaborative signals within limited context windows without incurring prohibitive latency.




\subsection{LLM-Generated Curation Rules}
\label{sec:generated-rules}

To efficiently curate the collaborative subgraph in Stage-R, $\text{LM}_{\text{Mem}}$ generates domain-specific heuristic rules in a zero-shot manner based on domain statistics.

Figure \ref{fig:llm_generated_curation_rules_1} presents the generated rules for the \textbf{Books} and \textbf{Goodreads} datasets, which emphasize content similarity (genre/theme) and social signals, respectively. Figure \ref{fig:llm_generated_curation_rules_2} shows the rules for the \textbf{MovieTV} and \textbf{Yelp} datasets, where recency and categorical constraints (e.g., cuisine/price) play a more dominant role. These interpretable rules act as a fast, domain-adaptive filter before memory synthesis. In our experiments, to ensure efficiency, we relied on statistical signals (e.g., recency, co-interactions) and used constant similarity scores instead of performing computationally expensive semantic calculations.

\subsection{Cost Estimation Methodology}
\label{sec:cost_methodology}

The cost estimates presented in Section \ref{sec:flexibility_analysis} and Figure \ref{fig:efficiency_bubble} are based on public cloud pricing for the Azure OpenAI Service Standard tier as of December 2025.\footnote{Official pricing source: \url{https://azure.microsoft.com/en-us/pricing/details/cognitive-services/openai-service/}. Pricing is subject to change.}


\begin{itemize}[leftmargin=*, itemsep=4pt]
    \item \textbf{High-Tier Model (gpt-4o):} \$2.50 per 1M input / \$10.00 per 1M output.
    \item \textbf{Low-Tier Models (gpt-4o-mini \& gpt-oss-120b):} \$0.15 per 1M input / \$0.60 per 1M output.
    \item \textbf{Local Deployment:} Negligible marginal cost.
\end{itemize}


\section{Detailed Related Works}
\label{app:detailed_related_work}

\subsection{Memory Architectures for LLM Agents}
Building autonomous agents capable of long-horizon tasks requires overcoming the inherent constraints of LLM context windows \cite{liu2024lost} and ensuring long-term knowledge retention. Early solutions combined LLMs with external vector databases \cite{johnson2019billion} to create Retrieval-Augmented Generation (RAG) pipelines \cite{lewis2020retrieval}. Recent advances like Graph RAG \cite{edge2024local} further demonstrate the value of structuring retrieved context into knowledge graphs for complex reasoning. Building on this, dedicated memory systems emerged. \textit{MemGPT} \cite{packer2023memgpt} introduced an OS-inspired virtual context management system, while \textit{Zep} \cite{rasmussen2025zep} structures memory into temporal knowledge graphs. Seminal works like \textit{Generative Agents} \cite{park2023generative} demonstrated how synthesizing high-level reflections from memory streams could drive believable agent behavior.

Complementarily, research explores learning-based memory policies \cite{xu2025amem, yan2025memory}, where a dedicated manager learns to optimize storage and retrieval for multi-hop reasoning.
General agent frameworks like LangChain~\cite{Chase_LangChain_2022} and AutoGPT~\cite{Significant_Gravitas_AutoGPT} have integrated modular components, such as tool use capabilities \cite{zhao2024let} and memory systems, to support complex workflows. Recent advancements also explore multi-dimensional memory management to handle long-context conversations~\cite{ye2026hmem}. While sophisticated, these systems are designed for general factual or conversational contexts, fundamentally neglecting the specialized, high-order connectivity required for graph-based collaborative domains. MemRec adopts the core principle of a decoupled memory manager ($\text{LM}_{\text{Mem}}$) but augments it with explicit graph context structure.

\subsection{Memory in Agentic RS}
The integration of memory into recommender systems has evolved from latent states in sequential models \cite{chen2022global, hidasi2015session, kang2018self} to explicit, dynamic structures managed by LLM agents.
Early applications of LLMs in recommendation explored stateless approaches, leveraging prompting or efficient architectures for ranking without maintaining persistent user states \cite{liu2023chatgpt, lyu2024llm, geng2022recommendation, ren2025easyrec, bao2023tallrec}. Subsequent works, such as \textit{Chat-REC} \cite{gao2023chat} and \textit{iAgent} \cite{xu2025iagent}, introduced explicit memory in the form of static user profiles or retrieved historical summaries, similar to standard RAG approaches. While enabling natural language interaction, these systems cannot adapt to evolving user interests based on real-time feedback.

To address plasticity, recent works introduce dynamic memory mechanisms, often incorporating planning or tool-using capabilities \cite{wang2024recmind, huang2025recommender, wang2024macrec, shu2023rah}. Systems like \textit{i$^2$Agent} \cite{xu2025iagent} and \textit{RecBot} \cite{tang2025interactive} employ a "self-reflection" mechanism, where the agent updates its own memory after an interaction. Similarly, simulation frameworks like \textit{AgentCF} \cite{zhang2024agentcf}, \textit{Agent4Rec} \cite{zhang2024generative}, and \textit{RecAgent} \cite{wang2025user} model users and items as agents with evolving memories to study emergent behaviors and feedback loops.

Crucially, these dynamic approaches remain bound to isolated, self-reflective memory.
The memory update is confined to the interacting user or item. While some recent works attempt to combine LLMs with graph structures for recommendation, they typically use LLMs for feature enhancement \cite{wei2024llmrec}, structure refinement \cite{wang-etal-2024-instructgraph}, graph vocabulary learning \cite{zhu2025llm}, or task-aware retrieval augmentation on dynamic graphs~\cite{tao2026task}, rather than for managing collaborative memory propagation in an agentic manner. 


Furthermore, our robustness analysis on niche users (Section \ref{sec:further_analysis}) connects agentic recommender systems to the broader challenge of user-side fairness and long-tail recommendation~\cite{chen2025investigating, chen2025leave}. While prior pre-LLM works typically address data sparsity for non-overlapping or long-tail users through cross-domain knowledge transfer~\cite{zhao2026double}, fairness-aware representation learning~\cite{chen2025causality}, or mining neutral signals and uncertainty from unlabeled data~\cite{ZCH24, zhao2025unlocking}, \texttt{MemRec} offers a novel agentic perspective. By using LLM-curated collaborative graphs, it democratizes recommendation quality, allowing users with sparse histories to fully benefit from community-level insights.

\section{Extended Efficiency and Modularity Analysis}
\label{sec:app_efficiency_analysis}


This appendix provides the detailed quantitative data supporting the analysis in Section \ref{sec:flexibility_analysis} of the main text.
Table \ref{tab:architectural_analysis_app} presents a comprehensive breakdown of different architectural configurations of \texttt{MemRec}. It reports performance metrics (H@1, N@5), alongside efficiency metrics including average experimental latency per user session, average total token consumption, and a qualitative cost estimate.

\begin{table*}[h!] 
\centering
\caption{Comprehensive architectural analysis across different model configurations. \textbf{Latency}: average online time measured in our experimental setup (sequential execution). \textbf{Tokens/U (On / Total)}: average tokens consumed per user during online inference versus the complete session (including asynchronous offline memory maintenance). \textbf{Cost}: qualitative estimate of online serving efficiency.}
\label{tab:architectural_analysis_app} 
\resizebox{\textwidth}{!}{%
\begin{tabular}{l|cc|cc|cc|c|l}
\toprule
& \multicolumn{2}{c|}{\textbf{Model Selection}} & \multicolumn{2}{c|}{\textbf{Performance}} & \multicolumn{3}{c|}{\textbf{Efficiency Metrics}} & \\
\textbf{Configuration} & $\text{LLM}_{\text{Rec}}$ & $\text{LM}_{\text{Mem}}$ & \textbf{H@1} & \textbf{N@5} & \textbf{Latency} & \textbf{Tokens/U (On / Total)} & \textbf{Cost (Online)} & \textbf{Key Takeaway} \\
\midrule
\textit{Vanilla LLM} & \texttt{4o-mini} & - & 0.330 & 0.524 & \textbf{$\sim$5.1s} & \textbf{$\sim$1.2k / $\sim$1.2k} & \textbf{Lowest} & \textit{No memory, fast but poor precision} \\
\textit{Single-User Mem} & \texttt{4o-mini} & \texttt{4o-mini (Iso.)} & 0.475 & 0.631 & $\sim$10.0s & $\sim$1.4k / $\sim$5.5k & Low & \textit{Memory overhead without collaboration} \\
\midrule
\rowcolor{gray!10}
\textbf{Standard} & \texttt{4o-mini} & \texttt{4o-mini} & 0.524 & 0.663 & $\sim$16.5s & \textbf{$\sim$1.5k / $\sim$6.8k} & Low & \textit{Collaborative gains over single-user} \\
\textbf{Vector} & \textit{Vector} & \texttt{4o-mini} & 0.209 & 0.387 & \underline{$\sim$5.3s} & \underline{- / $\sim$6.3k} & \underline{Lowest} & \textbf{Ultra-fast, pluggable reranker} \\
\textbf{Local-Qwen} & \texttt{4o-mini} & \texttt{Qwen-2.5-7B} & 0.470 & 0.627 & $\sim$34.0s$^\ddagger$ & $\sim$1.6k / $\sim$7.0k & Low$^\star$ & \textit{Best on-premise performance} \\
\textbf{Local-Llama} & \texttt{4o-mini} & \texttt{Llama-3-8B} & 0.360 & 0.550 & $\sim$34.4s$^\ddagger$ & $\sim$1.4k / $\sim$6.2k & Low$^\star$ & \textit{Alternative local option} \\
\midrule
\textbf{Ceiling} & \textbf{\texttt{gpt-4o}} & \texttt{4o-mini} & \textbf{0.580} & \textbf{0.722} & $\sim$10.4s & $\sim$1.5k / $\sim$6.9k & High & \textit{Peak performance, high serving cost} \\
\textbf{Cloud-OSS} & \texttt{4o-mini} & \textbf{\texttt{OSS-120B}} & \underline{0.561} & \underline{0.699} & $\sim$11.8s & $\sim$1.8k / $\sim$10.0k & \underline{Low} & \textbf{Near-ceiling results w/ moderate cost} \\
\bottomrule
\end{tabular}%
}
\footnotesize{
\\
$^\star$ The offline memory maintenance incurs zero API cost due to local hosting, leaving only the minimal online $\text{LLM}_{\text{Rec}}$ cost.
$^\ddagger$ Local latency is hardware-dependent (measured sequentially on single NVIDIA A5000 GPU) and not directly comparable to highly optimized cloud APIs.
}
\end{table*}

Table~\ref{tab:architectural_analysis_app} reports latencies measured in a sequential execution pipeline designed for rigorous benchmarking. In real-world deployments, user-perceived latency can be significantly reduced through standard engineered optimizations, such as aggressively \textit{caching} synthesized collaborative contexts (Stage-R outputs) for popular items to bypass redundant computations, and employing \textit{token streaming} for the final LLM output (Stage-ReRank) to drastically reduce the time-to-first-byte (TTFB) and improve perceived responsiveness.

The reported latencies require careful interpretation due to the nature of cloud API-based experimentation. Firstly, our experiments utilized standard, non-real-time API endpoints for all LLMs. Secondly, we observe a counter-intuitive result where the latency of the \textbf{Ceiling} configuration (\texttt{gpt-4o}) is lower than the \textbf{Standard} configuration (\texttt{4o-mini}), despite the former being a significantly larger model. We attribute this discrepancy to opaque operational factors related to the cloud provider's infrastructure, such as differential load balancing, resource allocation priorities, or transient network conditions at the time of measurement, rather than intrinsic differences in model inference speed. This highlights the variability inherent in benchmarking against black-box APIs.

\section{Additional Experimental Results}
\label{sec:appendix_additional_results}

\subsection{Results for Larger Candidate Set}
\label{sec:appendix_k20}
To demonstrate robustness, we present results for a larger candidate set ($N=20$) in Table \ref{tab:main_results_k20_1} and Table \ref{tab:main_results_k20_2}.

\begin{table*}[t]
\centering
\caption{Main results for \textbf{Books} and \textbf{Goodreads} (N=20). Notation follows Table~\ref{tab:main_results_1}; all improvements are significant ($p < 0.05$).}
\label{tab:main_results_k20_1}
\resizebox{\textwidth}{!}{%
\begin{tabular}{lrrrrrrrrrr} 
\toprule
& \multicolumn{5}{c}{\textbf{Books}} & \multicolumn{5}{c}{\textbf{Goodreads}} \\
\cmidrule(lr){2-6} \cmidrule(lr){7-11}
\textbf{Model} & \textbf{H@1} & \textbf{H@5} & \textbf{N@5} & \textbf{H@10} & \textbf{N@10} & \textbf{H@1} & \textbf{H@5} & \textbf{N@5} & \textbf{H@10} & \textbf{N@10} \\
\midrule
\addlinespace[0.3em] 
LightGCN & 0.1276 & 0.2622 & 0.1947 & 0.5512 & 0.2854 & 0.1617 & \underline{0.5566} & 0.3588 & \textbf{0.8177} & 0.4434 \\
SASRec & 0.0453 & 0.2353 & 0.1378 & 0.4896 & 0.2188 & 0.0699 & 0.3053 & 0.1859 & 0.5435 & 0.2621 \\
P5 & 0.1648 & 0.3051 & 0.2331 & 0.5216 & 0.3022 & 0.1038 & 0.2611 & 0.1798 & 0.5041 & 0.2572 \\
\midrule
\addlinespace[0.3em]
Vanilla LLM & 0.1730 & 0.4155 & 0.2955 & 0.6129 & 0.3599 & 0.0999 & 0.3245 & 0.2211 & 0.6712 & 0.3291 \\
iAgent & 0.3258 & 0.5069 & 0.4173 & 0.6209 & 0.4537 & 0.1621 & 0.4107 & 0.2871 & 0.6035 & 0.3490 \\
\midrule
\addlinespace[0.3em]
RecBot & 0.2471 & 0.4030 & 0.3247 & 0.5768 & 0.3801 & 0.1234 & 0.3364 & 0.2289 & 0.5583 & 0.2999 \\
AgentCF & 0.2470 & 0.5481 & 0.4026 & 0.7250 & 0.4594 & 0.1875 & 0.5427 & 0.3692 & 0.7805 & 0.4462 \\
i$^2$Agent & \underline{0.3712} & \underline{0.5947} & \underline{0.4874} & \underline{0.7387} & \underline{0.5336} & \underline{0.2065} & 0.5350 & \underline{0.3767} & 0.7428 & \underline{0.4435} \\
\midrule
\addlinespace[0.3em]
\rowcolor{gray!20}
MemRec (Ours) & \textbf{0.4236} & \textbf{0.6351} & \textbf{0.5332} & \textbf{0.7667} & \textbf{0.5756} & \textbf{0.2657} & \textbf{0.6062} & \textbf{0.4434} & \underline{0.7948} & \textbf{0.5042} \\
\textit{Improv.} & \multicolumn{1}{c}{\textit{+14.12\%}} & \multicolumn{1}{c}{\textit{+6.79\%}} & \multicolumn{1}{c}{\textit{+9.40\%}} & \multicolumn{1}{c}{\textit{+3.79\%}} & \multicolumn{1}{c}{\textit{+7.87\%}} & \multicolumn{1}{c}{\textit{+28.67\%}} & \multicolumn{1}{c}{\textit{+8.91\%}} & \multicolumn{1}{c}{\textit{+17.71\%}} & \multicolumn{1}{c}{-} & \multicolumn{1}{c}{\textit{+13.69\%}} \\
\bottomrule
\end{tabular}%
}
\end{table*}

\begin{table*}[t]
\centering
\caption{Main results for \textbf{MovieTV} and \textbf{Yelp} (N=20). Notation follows Table~\ref{tab:main_results_1}; all improvements are significant ($p < 0.05$).}
\label{tab:main_results_k20_2}
\resizebox{\textwidth}{!}{%
\begin{tabular}{lrrrrrrrrrr} 
\toprule
& \multicolumn{5}{c}{\textbf{MovieTV}} & \multicolumn{5}{c}{\textbf{Yelp}} \\
\cmidrule(lr){2-6} \cmidrule(lr){7-11}
\textbf{Model} & \textbf{H@1} & \textbf{H@5} & \textbf{N@5} & \textbf{H@10} & \textbf{N@10} & \textbf{H@1} & \textbf{H@5} & \textbf{N@5} & \textbf{H@10} & \textbf{N@10} \\
\midrule
\addlinespace[0.3em] 
LightGCN & 0.2657 & 0.5330 & 0.4064 & 0.6815 & 0.4537 & 0.2549 & 0.5437 & 0.4046 & 0.7481 & 0.4692 \\
SASRec & 0.2923 & 0.5128 & 0.4092 & 0.6311 & 0.4470 & 0.1678 & 0.3993 & 0.2879 & 0.5590 & 0.3389 \\
P5 & 0.1113 & 0.2769 & 0.1902 & 0.5137 & 0.2657 & 0.0634 & 0.2492 & 0.1537 & 0.5051 & 0.2354 \\
\midrule
\addlinespace[0.3em]
Vanilla LLM & 0.2379 & 0.5003 & 0.3648 & 0.7261 & 0.4406 & 0.0254 & 0.1461 & 0.0831 & 0.5128 & 0.2010\\
iAgent & 0.3236 & 0.5362 & 0.4331 & 0.6762 & 0.4778 & 0.3236 & 0.5658 & 0.4499 & 0.6597 & 0.4799 \\
\midrule
\addlinespace[0.3em]
RecBot & 0.2420 & 0.4201 & 0.3316 & 0.6015 & 0.3895 & 0.1949 & 0.3742 & 0.2851 & 0.5519 & 0.3414 \\
AgentCF & 0.2870 & 0.6288 & 0.4648 & 0.7616 & 0.5077 & 0.1115 & 0.3897 & 0.2512 & 0.6372 & 0.3309 \\
i$^2$Agent & \underline{0.3822} & \underline{0.6367} & \underline{0.5178} & \underline{0.7735} & \underline{0.5617} & \underline{0.3287} & \underline{0.6083} & \underline{0.4744} & \underline{0.7562}& \underline{0.5216} \\
\midrule
\addlinespace[0.3em]
\rowcolor{gray!20}
MemRec (Ours) & \textbf{0.4750} & \textbf{0.7543} & \textbf{0.6212} & \textbf{0.8752} & \textbf{0.6606} & \textbf{0.3620} & \textbf{0.6329} & \textbf{0.5035} & \textbf{0.7708} & \textbf{0.5478} \\
\textit{Improv.} & \multicolumn{1}{c}{\textit{+24.28\%}} & \multicolumn{1}{c}{\textit{+18.47\%}} & \multicolumn{1}{c}{\textit{+19.97\%}} & \multicolumn{1}{c}{\textit{+13.15\%}} & \multicolumn{1}{c}{\textit{+17.61\%}} & \multicolumn{1}{c}{\textit{+10.13\%}} & \multicolumn{1}{c}{\textit{+4.04\%}} & \multicolumn{1}{c}{\textit{+6.13\%}} & \multicolumn{1}{c}{\textit{+1.93\%}} & \multicolumn{1}{c}{\textit{+5.02\%}} \\
\bottomrule
\end{tabular}%
}
\end{table*}

\subsection{Rationale Quality Analysis}
\label{sec:app_rationale_analysis}

To assess the qualitative impact of collaborative memory on reasoning output, we conducted a human-aligned evaluation using GPT-4o as an automated judge on the \texttt{books} subset. The detailed evaluation protocol, including the model mapping and the exact prompts used for the GPT-4o judge, is described in Appendix \ref{sec:app_rationale_eval_protocol}.

We compared rationales generated by three model configurations: \textbf{Base LLM} (Vanilla, no memory), \textbf{MemRec w/o Collab} (static user history only), and \textbf{MemRec (Full)} (collaborative memory). GPT-4o rated each rationale on a Likert scale (1-5) across three distinct dimensions: \textit{Specificity} (richness of item details), \textit{Relevance} (connection to user interests), and \textit{Factuality} (accuracy of claims).

Figure \ref{fig:rationale_eval} presents the average scores with 95\% confidence intervals and significance annotations (paired t-test). The results reveal distinct trends regarding the role of different memory types:

\begin{itemize}[leftmargin=*, itemsep=2pt, topsep=3pt]
    \item \textbf{Specificity exhibits a clear step-wise improvement.} Adding user history (w/o Collab) significantly improves specificity over the Base LLM ($p<0.001$), likely by grounding the generation in the user's genre. Crucially, adding collaborative memory (Full) yields a further significant boost ($p<0.001$). This confirms that neighbor signals provide the rich, specific item details needed for high-quality justification that user history alone cannot provide.

    \item \textbf{Collaborative signals are key to perceived Relevance.} Surprisingly, user history alone (\textit{w/o Collab}) did not yield a statistically significant improvement over the Base LLM in perceived relevance ($p>0.05$). However, the full collaborative context (\textit{MemRec}) achieved a substantial and significant increase ($p<0.001$). This suggests that simply mentioning user history is insufficient; grounding recommendations in peer experiences makes them feel significantly more relevant and convincing to the judge.

    \item \textbf{MemRec maintains high Factuality.} While all models maintain high factuality scores (>4.0), \texttt{MemRec} achieves a slight but statistically significant improvement over the others ($p<0.001$ vs w/o Collab), indicating that grounding generation based on real collaborative memories helps reduce hallucinations compared to ungrounded generation.
\end{itemize}

\subsection{Latency and Token Breakdown Analysis}
\label{sec:appendix_token_analysis}

A critical aspect of \texttt{MemRec}'s cost-efficiency lies in how it utilizes tokens relative to standard commercial pricing structures.

Most commercial LLM providers adopt an \textbf{asymmetric pricing model}, where output (generated) tokens are significantly more expensive than input (context) tokens (typically a 3x to 4x ratio, see Appendix \ref{sec:cost_methodology}). \texttt{MemRec}'s architecture is inherently designed to exploit this structure.

As shown in Table \ref{tab:stage_token_breakdown}, our key memory operations, Stage-R (Synthesis) and Stage-W (Propagation), are heavily \textit{input-biased}. They digest large volumes of raw collaborative context (cheap input) to produce highly condensed, structured insights (expensive output). For example, in the Standard configuration, input tokens account for \textbf{nearly 80\%} of the total usage. This makes the effective cost of running \texttt{MemRec} significantly lower than a naive estimation based on total token counts would suggest.

\begin{table*}[h!]
\centering
\caption{Detailed breakdown of average Input vs. Output token consumption per stage per user (measured on \texttt{books} for the Standard configuration). The high Input/Output ratio in memory stages exploits the asymmetric pricing of commercial LLMs.}
\label{tab:stage_token_breakdown}
\begin{tabular}{l|c|rr|r|c}
\toprule
\textbf{Pipeline Stage} & \textbf{Primary Role} & \textbf{Avg Input} & \textbf{Avg Output} & \textbf{Total} & \textbf{I/O Ratio} \\
\midrule
Stage-R (Synthesis) & $\text{LM}_{\text{Mem}}$ & $\sim$1,400 & $\sim$460 & $\sim$1,860 & 3.0 : 1 \\
Stage-ReRank & $\text{LLM}_{\text{Rec}}$ & $\sim$1,100 & $\sim$450$^\dagger$ & $\sim$1,550 & 2.4 : 1 \\
Stage-W (Async) & $\text{LM}_{\text{Mem}}$ & $\sim$1,600 & $\sim$315 & $\sim$1,915 & 5.1 : 1 \\
\midrule
\textbf{Total per User} & - & \textbf{$\sim$5,100} & \textbf{$\sim$1,300} & \textbf{$\sim$6,400} & \textbf{3.9 : 1} \\
\bottomrule
\end{tabular}%
\footnotesize{\\$^\dagger$ A significant portion of Stage-ReRank output is dedicated to generating interpretable rationales, adding user value.}
\end{table*}

\subsection{Hyperparameter Analysis}
\label{sec:appendix_hyper}

\begin{figure*}[h!] 
    \centering
    \begin{subfigure}[b]{0.48\textwidth}
        \centering
        \includegraphics[width=\textwidth]{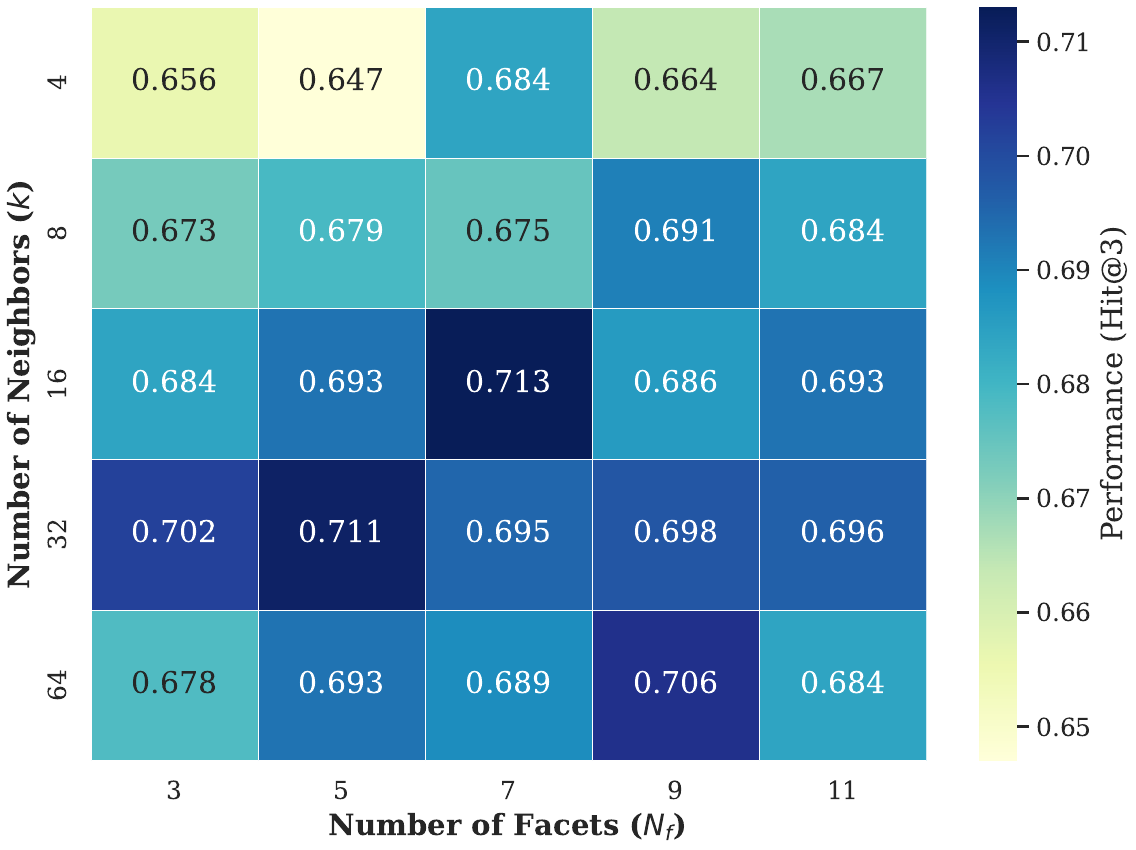}
        \caption{Hit@3}
        \label{fig:hyper_h3}
    \end{subfigure}
    \hfill 
    \begin{subfigure}[b]{0.48\textwidth}
        \centering
        \includegraphics[width=\textwidth]{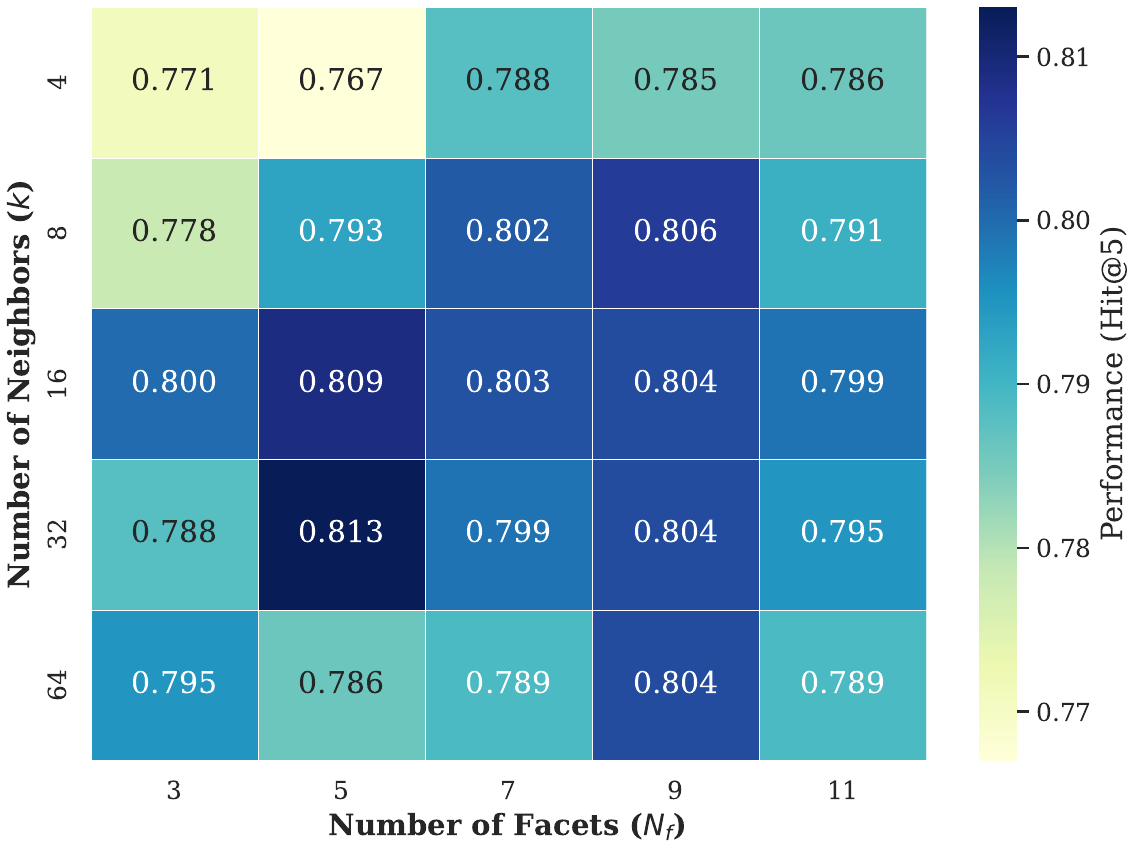}
        \caption{Hit@5}
        \label{fig:hyper_h5}
    \end{subfigure}
    
    \vspace{1em} 

    \begin{subfigure}[b]{0.48\textwidth}
        \centering
        \includegraphics[width=\textwidth]{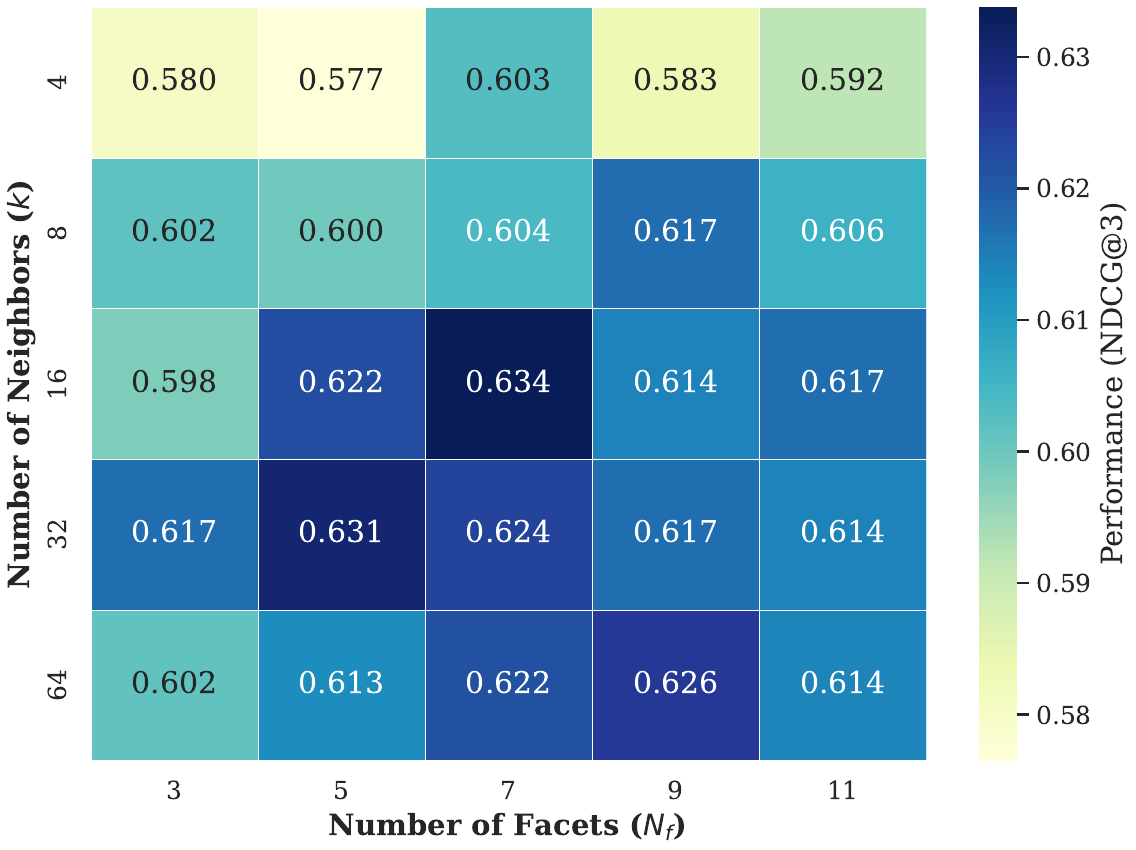}
        \caption{NDCG@3}
        \label{fig:hyper_n3}
    \end{subfigure}
    \hfill 
    \begin{subfigure}[b]{0.48\textwidth}
        \centering
        \includegraphics[width=\textwidth]{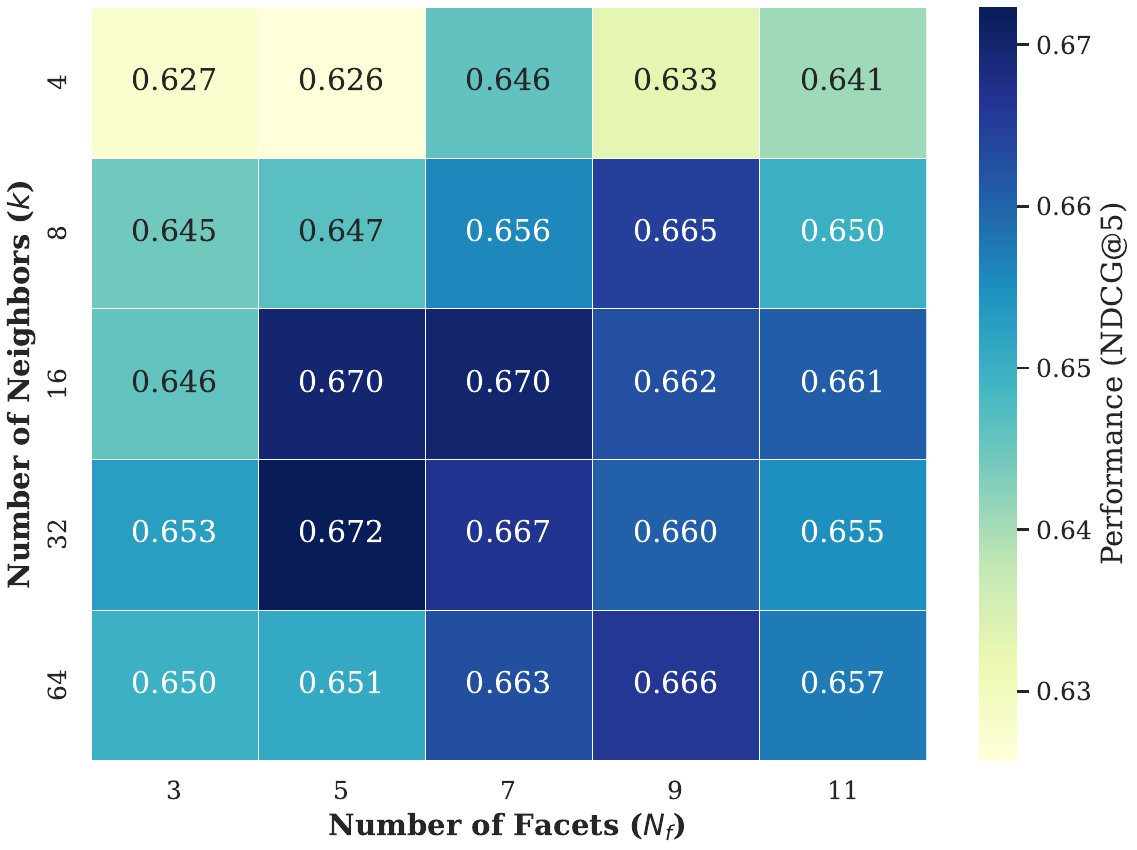}
        \caption{NDCG@5}
        \label{fig:hyper_n5}
    \end{subfigure}
    
    \caption{Hyperparameter sensitivity analysis for additional metrics on the \texttt{books} subset. Trends are consistent with H@1 shown in the main text.}
    \label{fig:hyperparams-appendix-full}
\end{figure*}

We analyze the sensitivity of \texttt{MemRec} to its two main hyperparameters on the \texttt{books-1k} subset: the number of neighbors $k$ (Stage-R Curation) and the number of facets $N_f$ (Stage-R Synthesis). While the primary metric H@1 is shown in Figure \ref{fig:hyperparam_heatmap_h1} in the main text, Figure \ref{fig:hyperparams-appendix-full} presents the heatmaps for additional metrics (H@3, H@5, NDCG@3, NDCG@5). The trends across these metrics are consistent with H@1, confirming the robustness of the optimal hyperparameter region.

\subsection{Full Metrics for Architectural Analysis}
\label{sec:app_additional_results_arch}
Table \ref{tab:app_arch_full_metrics} presents the comprehensive performance metrics covering Hit Rate (H@K) and NDCG (N@K) at varying cutoff points ($K=\{1, 3, 5\}$) across three datasets. These detailed results substantiate the findings discussed in Section \ref{sec:architectural_analysis}, demonstrating the consistent superiority of MemRec over both the Vanilla baseline and the Naive Agent across diverse ranking depths.

\begin{table*}[h]
\centering
\caption{Comprehensive performance comparison for the architectural analysis across datasets.  \textbf{Vanilla LLM} serves as the non-memory baseline. \textbf{Naive Agent} utilizes raw, uncurated collaborative context. \textbf{MemRec} employs the proposed decoupled architecture. The best performance in each metric per dataset is marked in \textbf{bold}.}
\label{tab:app_arch_full_metrics}
\begin{tabular}{llccccc}
\toprule
 &  & \multicolumn{3}{c}{\textbf{Hit Rate @ K}} & \multicolumn{2}{c}{\textbf{NDCG @ K}} \\
\cmidrule(lr){3-5} \cmidrule(lr){6-7}
\textbf{Dataset} & \textbf{Model} & \textbf{H@1} & \textbf{H@3} & \textbf{H@5} & \textbf{N@3} & \textbf{N@5} \\
\midrule
\multirow{3}{*}{\textbf{Books}} 
& Vanilla LLM & 0.330 & 0.549 & 0.724 & 0.450 & 0.524 \\
& Naive Agent & 0.390 & 0.638 & 0.760 & 0.533 & 0.583 \\
& \textbf{MemRec} & \textbf{0.524} & \textbf{0.700} & \textbf{0.795} & \textbf{0.625} & \textbf{0.663} \\
\midrule
\multirow{3}{*}{\textbf{Yelp}} 
& Vanilla LLM & 0.176 & 0.522 & 0.684 & 0.370 & 0.437 \\
& Naive Agent & 0.242 & 0.486 & 0.639 & 0.383 & 0.446 \\
& \textbf{MemRec} & \textbf{0.489} & \textbf{0.686} & \textbf{0.792} & \textbf{0.604} & \textbf{0.648} \\
\midrule
\multirow{3}{*}{\textbf{MovieTV}} 
& Vanilla LLM & 0.407 & 0.774 & 0.861 & 0.610 & 0.646 \\
& Naive Agent & 0.418 & 0.605 & 0.795 & 0.523 & 0.602 \\
& \textbf{MemRec} & \textbf{0.563} & \textbf{0.791} & \textbf{0.894} & \textbf{0.696} & \textbf{0.738} \\
\bottomrule
\end{tabular}%
\end{table*}

\section{Qualitative Case Study: A Complete Collaborative Journey}
\label{sec:appendix_case_study}

This case study illustrates the complete workflow of \texttt{MemRec} for \textbf{User 2057, a fan of Young Adult (YA) fantasy and graphic novels.} Figure \ref{fig:complete_case_study} demonstrates how collaborative signals are synthesized (Stage-R), leveraged for reasoning (Stage-ReRank), and propagated back into the memory graph (Stage-W). For brevity and clarity, we display only representative subsets of retrieved neighbors (Stage-R) and propagated updates (Stage-W) to illustrate the process. We use \collab{blue text} to highlight collaborative signals and \user{orange text} to indicate user-specific signals.

\section{Prompt Templates and Contexts}
\label{sec:app_prompts}


This appendix provides the complete prompt templates governing the memory management ($\text{LM}_{\text{Mem}}$) and reasoning ($\text{LLM}_{\text{Rec}}$) agents, ensuring full transparency and reproducibility of our framework.

\subsection{Meta-Prompt Template}
\label{sec:prompts}


To enable zero-shot domain adaptation for Stage-R neighbor pruning, we employ a Meta-Prompt Template (Figure~\ref{fig:meta_prompt_template}). This high-level instruction inputs structured domain descriptions (metadata, statistics) to guide $\text{LM}_{\text{Mem}}$ in synthesizing interpretable, domain-specific heuristic rules offline.

\subsection{Domain-Specific Prompt Contexts}
\label{sec:domain-prompts}


Figure~\ref{fig:domain_context} presents the specific Domain Context Blocks injected into the Meta-Prompt for our four evaluated datasets. These blocks outline key metadata fields, interaction modes, and unique domain characteristics to generate effective, tailored rules offline.

\subsection{Stage-R Memory Synthesis Prompt}
\label{sec:prompt_stage_r_synthesis}


The \textbf{Stage-R Synthesis Prompt} (Figure~\ref{fig:prompt_synthesis}) enables $\text{LM}_{\text{Mem}}$ to distill raw neighbor memories into high-signal "memory facets." It instructs the LLM to identify common themes from the user's and neighbors' memories, synthesizing them into structured facets with confidence scores to form the collaborative context $(M_{\text{collab}})$ for downstream reasoning.

\subsection{Stage-ReRank Scoring Prompt}
\label{sec:prompt_rerank}

The final ranking decision in Stage-ReRank is performed by the reasoning agent ($\text{LLM}_{\text{Rec}}$) using the \textbf{Stage-ReRank Scoring Prompt}.
As displayed in Figure \ref{fig:prompt_rerank}, this prompt is designed to ground the LLM's reasoning in the provided context. It integrates three key inputs: the user's natural language instruction, specific details of the candidate item being evaluated, and, crucially, the synthesized collaborative memory facets ($M_{\text{collab}}$). The prompt instructs the LLM to synthesize these signals to generate a calibrated relevance score (between 0 and 1) along with a concise, natural language rationale justifying the score based on collaborative evidence.

\subsection{Stage-W Propagation Prompts}\label{sec:prompt_stage_w}Following a user interaction (e.g., clicking an item), the \textbf{Stage-W Propagation Prompts} are employed asynchronously by $\text{LM}_{\text{Mem}}$ to evolve the semantic graph. While efficiently implemented as a single LLM call to minimize latency, we conceptually decompose this process into two distinct logical operations, as illustrated in Figure \ref{fig:prompt_propagation}:\begin{enumerate} \item \textbf{User/Item Memory Update}: Updating the narrative memories of the interacting user and item nodes to reflect the new interaction.\item \textbf{Collaborative Propagation}: Identifying relevant neighboring nodes and propagating insights from the interaction to update their memories, thereby enriching the graph's collaborative signals for future retrievals.\end{enumerate}Figure \ref{fig:prompt_propagation} shows the comprehensive prompt that handles these updates concurrently.

\begin{table*}[h]
\centering
\caption{Design comparison of neighbor curation approaches. Our approach uniquely balances the interpretability and efficiency of rule-based methods with the domain-adaptivity of learning-based methods, in a zero-shot manner.}
\label{tab:curation_comparison}
\begin{tabularx}{\linewidth}{lXXX}
\toprule
& \textbf{Rule-based} \newline {\small (Traditional heuristic)} & \textbf{LLM-generated rules} \newline {\small (\textbf{Our approach})} & \textbf{Learned $f_\theta$} \newline {\small (Neural scorer)} \\
\midrule
Training required? & No & \textbf{No} & Yes \\
Domain-adaptive? & Limited & \textbf{Yes} & \textbf{Yes} \\
Interpretable? & \textbf{High} & \textbf{High} & Low \\
Inference cost & \textbf{$<$1ms} & \textbf{$<$1ms}\textsuperscript{a} & \textbf{$<$1ms} \\
Performance (est.) & Moderate & Good & \textbf{Best} \\
Generalization & \textbf{High} & \textbf{High} & Low \\
Implementation & \textbf{Simple} & Moderate & Complex \\
\bottomrule
\end{tabularx}
\footnotesize
\textsuperscript{a} The cost reflects applying pre-generated rules online. The one-time offline rule generation by the LLM is negligible per inference.
\end{table*}

\begin{table*}[h]
\centering
\caption{Neighbor Filtering Precision (evaluated on a 1,000-user test subset with $k=16$). The proposed LLM curation drastically reduces noisy item neighbors compared to a generic heuristic rule.}
\label{tab:filtering_precision}
\begin{tabular}{lccc}
\toprule
\textbf{Metric} & \textbf{Generic Rule (w/o LLM)} & \textbf{LLM Curation (Ours)} & \textbf{$\Delta$} \\ 
\midrule
Total Neighbors Processed & 15,782 & 15,782 & - \\
Irrelevant Items Retained & 1,188 & 311 & \textbf{-73.8\%} $\downarrow$ \\
Irrelevant Users Retained & 1,754 & 1,866 & +6.4\% $\uparrow$ \\
Overall Filtering Precision & 0.813 & \textbf{0.861} & \textbf{+5.9\%} $\uparrow$ \\ 
\bottomrule
\end{tabular}
\end{table*}


\subsection{Rationale Quality Evaluation Protocol}
\label{sec:app_rationale_eval_protocol}

We evaluate the quality of the generated explanations (rationales) using GPT-4o as an automated judge. The judge is instructed to score rationales from three different models independently based on Specificity, Relevance, and Factuality on a 1-5 Likert scale.

The three models evaluated are mapped as follows:
\begin{itemize}[noitemsep, topsep=0pt]
    \item \textbf{Model A}: Base LLM (Vanilla LLM)
    \item \textbf{Model B}: MemRec w/o Collab (Isolated Memory)
    \item \textbf{Model C}: \texttt{MemRec} (Full)
\end{itemize}

The system prompt providing the evaluation criteria and scoring rubric, along with the user prompt template used for the GPT-4o judge, are presented in Figure \ref{fig:app_prompt_rationale_eval}. We use `gpt-4o` with a temperature of 0 to ensure consistent evaluation.

\section{Methodology Analysis}
\label{sec:app_methodology_analysis}

\subsection{Comparison of Curation Approaches}
\label{sec:curation_comparison}

To justify our design choice for the neighbor pruning step in Stage-R, Table \ref{tab:curation_comparison} presents a comparative analysis of three distinct approaches: traditional rule-based heuristics (e.g., Random Walk~\cite{perozzi2014deepwalk}), our proposed LLM-generated rules, and a fully learned neural scorer ($f_\theta$, e.g., GNN-based attention weights~\cite{velickovic2018graph}). Our \textbf{LLM-generated rules} approach occupies a sweet spot. Unlike traditional heuristics, it is \textit{domain-adaptive} (rules are tailored to specific dataset statistics) and offers better estimated performance. Unlike a learned scorer, it requires \textit{no training data}, maintains \textit{high interpretability} (rules are human-readable), and ensures extremely low online inference cost. This balance makes it a practical and robust solution for efficiently curating high-signal subgraphs.

\subsection{Quantitative Analysis of LLM Curation}

To further demonstrate the superiority of the LLM-guided context curation over generic heuristics, we conducted a quantitative precision analysis. Recognizing the absence of an absolute "gold standard" for neighbor relevance in this unsupervised context, we established an intuitive statistical proxy to identify highly probable irrelevant neighbors across a 1,000-user test subset (processing 15,782 neighbors in total). Specifically, we flagged:

\begin{itemize}
    \item \textbf{Irrelevant Items (Outdated):} Items with a recency score $< 0.3$ (corresponding to interactions older than approximately 255 days). An item a user interacted with nearly a year ago typically no longer reflects their current, actively evolving preferences.
    \item \textbf{Irrelevant Users (Low Similarity):} Users with an interaction overlap ratio $< 0.1$. For instance, a peer who shares almost zero common interaction history with the target user provides exceptionally weak traditional collaborative signals.
\end{itemize}

We compared our zero-shot LLM curation against a "Generic Rule" variant (w/o LLM Curation), which utilizes a standard, non-adaptive heuristic applying equal weights to all available normalized features (e.g., recency, co-interaction count, metadata overlap) without domain-specific semantic adjustments.


As shown in Table \ref{tab:filtering_precision}, our zero-shot LLM curation provides a significant advantage, particularly in filtering noisy items. The LLM curation effectively reduced irrelevant item neighbors by 73.8\%, acting as a critical noise buffer for the downstream reasoning agent. This capability to identify and filter noisy interactions shares conceptual goals with hardness-adaptive negative sampling in traditional representation learning \cite{lai2026matryoshka}. However, rather than relying on complex training objectives, \texttt{MemRec} achieves robust noise reduction through zero-shot semantic curation. Interestingly, the LLM curation retained slightly more users (+6.4\%) with low ID-overlap. We view this as a feature rather than a flaw: the LLM identifies semantic similarity from memory narratives (e.g., two users who both enjoy "dystopian YA novels" even if they have not clicked the exact same items), thereby prioritizing diverse but conceptually relevant peers over rigid statistical overlap.

\clearpage

\begin{figure*}[h!]
\centering
\footnotesize 

\begin{subfigure}[b]{\textwidth}
\begin{promptbox}[Stage-R: Collaborative Synthesis ($\text{LM}_{\text{Mem}}$)]
\setlength{\baselineskip}{1.1em} 
\textbf{Task:} Synthesize collaborative context for User 2057 based on retrieved neighbors (showing 3 of $k=16$).

\textbf{Selected Neighbor Memories (Input Noise):}
\begin{itemize}[noitemsep, topsep=0pt, leftmargin=*]
    \item \textbf{Item Neighbor (*Pastworld*):} "...In 2050, civilization has become sterile, controlled... boring. A dystopian setting..."
    \item \textbf{Item Neighbor (*The Lighthouse Land*):} "...Thirteen-year-old Jamie... leaves New York... for an island off the coast... magical adventure."
    \item \textbf{User Neighbor (User-4023):} "Recent: *PrettyTOUGH*, *Head Games*, *Heartbreak River*." (Interactions implying YA contemporary themes).
    
    \vspace{3pt}
    \footnotesize\textit{... (processing a total of 16 neighbors' contextual representations)}
    \vspace{3pt}
\end{itemize}

\vspace{4pt} \hrule \vspace{4pt}

\textbf{Synthesized Collaborative Context (High-Signal Output):}
The community around User 2057 engages deeply with speculative narratives. Key themes include \collab{dystopian and speculative fiction}, often featuring young protagonists in \collab{adventurous themes}. There is a strong appreciation for \collab{fantasy novels with strong world-building} and complex, often morally ambiguous character dynamics.

\textbf{Extracted Collaborative Facets:}
\begin{itemize}[noitemsep, topsep=0pt, leftmargin=*]
    \item \textbf{Facet 1:} \collab{Interest in dystopian and speculative fiction.}
    \item \textbf{Facet 2:} \collab{Preference for young adult literature with adventurous themes.}
    \item \textbf{Facet 3:} \collab{Liking for fantasy novels with strong world-building.}
\end{itemize}
\rat{Rationale: "$\text{LM}_{\text{Mem}}$ successfully distills common themes from YA fantasy neighbors into compact facets (highlighted in \collab{blue}), capturing the user's affinity for expansive, speculative worlds."}
\end{promptbox}
\end{subfigure}

\vspace{6pt} 

\begin{subfigure}[b]{\textwidth}
\begin{promptbox}[Stage-ReRank: Grounded Reasoning ($\text{LLM}_{\text{Rec}}$)]
\setlength{\baselineskip}{1.1em}
\textbf{User Instruction:} "\user{I seek a book that can provide a visually immersive experience... stunning visuals and a captivating storyline... as detailed and engaging as a well-crafted piece of code...}" (User explicitly requests a \textbf{graphic novel}).

\textbf{Input Context:} User 2057's Profile (YA Fantasy history) + \textbf{Synthesized Collaborative Context} (from Stage-R).

\vspace{4pt} \hrule \vspace{4pt}

\textbf{Recommended Item:} \textit{Attack on Titan: No Regrets Complete Color Edition}

\textbf{Generated Rationale:}
"\textit{'Attack on Titan: No Regrets'} is a \user{graphic novel} known for its \user{stunning visuals} and \collab{complex storyline}, aligning perfectly with your request for an immersive narrative. It strongly resonates with the \collab{collaborative insight} that users similar to you enjoy \collab{dystopian and speculative fiction} with \collab{strong world-building} and emotionally layered storytelling."

\end{promptbox}
\end{subfigure}

\vspace{6pt} 

\begin{subfigure}[b]{\textwidth}
\begin{promptbox}[Stage-W: Asynchronous Propagation ($\text{LM}_{\text{Mem}}$)]
\setlength{\baselineskip}{1.1em}
\textbf{Event:} User 2057 interacts positively with \textit{Attack on Titan}.

\vspace{4pt} \hrule \vspace{4pt}

\textbf{1. User Memory Update (User 2057):}
\begin{itemize}[noitemsep, topsep=0pt, leftmargin=*]
    \item \textbf{[OLD]}: "Fan of YA fantasy, adventurous themes, and strong world-building."
    \item \textbf{[NEW]}: "Fan of YA fantasy and \collab{dystopian fiction}. Confirmed strong interest in the \user{'graphic novel format'} combined with \collab{complex, emotional storytelling}."
\end{itemize}

\textbf{2. Target Item Memory Update (\textit{Attack on Titan}):}
\begin{itemize}[noitemsep, topsep=0pt, leftmargin=*]
    \item \textbf{[OLD]}: "...graphic novel, spin-off series..."
    \item \textbf{[NEW]}: "...graphic novel known for stunning visuals. \collab{Validated appeal for YA fantasy readers seeking immersive dystopian worlds and complex narratives.}"
\end{itemize}
\rat{Rationale: "The item's memory is updated to reflect its appeal to the specific user profile that engaged with it."}

\textbf{3. Neighbor Propagation Update (User-4023):}
\begin{itemize}[noitemsep, topsep=0pt, leftmargin=*]
    \item \textbf{[OLD]}: "Interested in humorous young adult novels... explores teenage experiences..."
    \item \textbf{[NEW]}: "...appreciates \collab{strong character development and emotional depth}. May enjoy narratives that blend humor with serious, complex themes."
\end{itemize}
\rat{Rationale: "Insights on 'strong character development' from the interaction are propagated to relevant neighbors, enriching the collaborative graph."}
\end{promptbox}
\end{subfigure}

\caption{\textbf{Complete Collaborative Journey (User 2057).} The figure illustrates the data flow across MemRec's three stages. \textbf{Stage-R:} $\text{LM}_{\text{Mem}}$ synthesizes \collab{collaborative signals (blue)} from noisy neighbors (e.g., dystopian, YA fantasy themes). \textbf{Stage-ReRank:} $\text{LLM}_{\text{Rec}}$ combines these signals with the user's explicit intent for a \user{graphic novel with stunning visuals (orange)} to recommend \textit{Attack on Titan}. \textbf{Stage-W:} Following interaction, the validated insights are propagated back, updating the user, the item, and relevant neighbors like User-4023. Note that only representative subsets shown for brevity.}
\label{fig:complete_case_study}
\end{figure*}


\clearpage
\begin{figure*}[h!]
\begin{promptbox}[Meta-Prompt Template for Rule Generation]
You are an expert AI engineer specializing in recommender systems and graph-based memory networks.
Your task is to generate a set of domain-specific heuristic rules for a \textbf{collaborative neighbor pruning} algorithm. The goal of this algorithm is to select the top-k most relevant neighbors (users or items) from a candidate graph to build a compact, high-signal context for a downstream LLM recommender (MemRec).

Here is the context for the specific recommendation domain:

\noindent\rule{\textwidth}{0.4pt}
\textbf{DOMAIN CONTEXT}

\begin{itemize}[leftmargin=*, nosep]
\item \textbf{Domain Name:} \textcolor{promptkeyword}{\{Domain Name\}}
\item \textbf{Primary Interaction:} \textcolor{promptkeyword}{\{Primary Interaction with example\}}
\item \textbf{Key Metadata:} \textcolor{promptkeyword}{\{Key Metadata\}}
\item \textbf{Available Features:}
\begin{itemize}
\item \texttt{edge\_weight}: \textcolor{promptkeyword}{\{Domain-specific explanation\}}
\item \texttt{recency\_days}: \textcolor{promptkeyword}{\{Domain-specific explanation\}}
\item \texttt{co\_interaction\_count}: \textcolor{promptkeyword}{\{Domain-specific explanation\}}
\item \texttt{metadata\_overlap\_score}: \textcolor{promptkeyword}{\{Domain-specific explanation\}}\item \texttt{memory\_similarity\_score}: \textcolor{promptkeyword}{\{Domain-specific explanation\}}
\end{itemize}

\end{itemize}
\noindent\rule{\textwidth}{0.4pt}

\textbf{INSTRUCTIONS:}

1. Based \textit{only} on the domain context provided, generate 3-5 high-priority, interpretable ranking rules.

2. The rules should explain how to \textit{combine} or \textit{prioritize} the available features to find the best neighbors for \textit{this specific domain}.

3. Be specific about thresholds and weights. For example:

\;\;\;\;\;\;- Good: "Prioritize users with `co\_interaction\_count` > 3 AND apply a 2.0x multiplier to `metadata\_overlap\_score`"

\;\;\;\;\;\;- Bad: "Use metadata when relevant"

4. Consider that book recommendations are highly content-driven (genre, author, themes) and users often have stable long-term preferences.

\noindent\rule{\textwidth}{0.4pt}
\textbf{OUTPUT FORMAT:}

Rule 1: [Your rule here]

Rule 2: [Your rule here]

Rule 3: [Your rule here]

...

\end{promptbox}
\caption{The generic meta-prompt template used by $\text{LM}_{\text{Mem}}$ to generate domain-specific curation rules.}
\label{fig:meta_prompt_template}
\end{figure*}

\clearpage

\begin{figure*}[h!]
\begin{promptbox}[Context: InstructRec-Books]
\textbf{Domain Name:} InstructRec-Books \\
\textbf{Primary Interaction:} Explicit ratings with text-based preference instructions. Example: "I love fantasy novels with strong female protagonists" \\
\textbf{Key Metadata:} \texttt{title}, \texttt{description} (genre hints, author info) \\
\textbf{Domain Characteristics:} Content-driven, stable preferences, sparse interactions.
\end{promptbox}

\begin{promptbox}[Context: InstructRec-GoodReads]
\textbf{Domain Name:} InstructRec-GoodReads \\
\textbf{Primary Interaction:} Explicit ratings in a social reading context. \\
\textbf{Key Metadata:} \texttt{title} (series info), \texttt{description}. \\
\textbf{Domain Characteristics:} Very dense graph (avg 52.7 books/user), strong community effects, series-aware reading.
\end{promptbox}

\begin{promptbox}[Context: InstructRec-MovieTV]
\textbf{Domain Name:} InstructRec-MovieTV \\
\textbf{Primary Interaction:} Explicit ratings with viewing preferences. \\
\textbf{Key Metadata:} \texttt{title}, \texttt{description} (Plot, Cast). \\
\textbf{Domain Characteristics:} Sparse graph, recency matters (trending content), volatile preferences.
\end{promptbox}

\begin{promptbox}[Context: InstructRec-Yelp]
\textbf{Domain Name:} InstructRec-Yelp \\
\textbf{Key Metadata:} \texttt{categories} (Cuisine), \texttt{attributes} (Price, WiFi). \\
\textbf{Domain Characteristics:} Context-rich but sparse. Strong categorical constraints (cuisine/price/location). Recency is critical.
\end{promptbox}
\caption{The specific `DOMAIN CONTEXT` blocks injected into the meta-prompt for each dataset.}
\label{fig:domain_context}
\end{figure*}

\clearpage
\begin{figure*}[h!]
\begin{promptbox}[Generated Rules: Books (Content-Driven)]
\textbf{Rule 1: Content Similarity Boost}

- If \feat{metadata\_overlap\_score} $> 0.6$ (strong genre/theme match): \textbf{Apply 2.5x multiplier} to base score.

- \rat{Rationale: Books are highly content-driven; genre/author similarity is the strongest signal.}

\textbf{Rule 2: Collaborative Filtering with Threshold}

- If \feat{co\_interaction\_count} $> 3$: \textbf{Apply 1.8x multiplier} to \feat{edge\_weight}.

- Additional 1.5x boost if \feat{memory\_similarity\_score} $> 0.5$.

- \rat{Rationale: Meaningful CF signal requires sufficient overlap (>3 books).}

\textbf{Rule 3: Mild Recency Decay}

- If \feat{recency\_days} $> 180$: Apply decay factor: $\exp(-0.004 \times \feat{recency\_days})$.

- Else: No decay.

- \rat{Rationale: Books have long-term appeal; old interactions remain relevant.}

\textbf{Rule 4: Memory-Enhanced Ranking}

- For item neighbors: Boost by $1.0 + 1.2 \times \feat{memory\_similarity\_score}$.

- For user neighbors: Boost by $1.0 + 0.8 \times \feat{memory\_similarity\_score}$.

- \rat{Rationale: User memory captures learned preferences; stronger for items.}
\end{promptbox}

\begin{promptbox}[Generated Rules: GoodReads (Social/Dense)]
\textbf{Rule 1: High Co-interaction Boost}

- If \feat{co\_interaction\_count} $> 10$: \textbf{Apply 2.0x multiplier} to \feat{edge\_weight}.

- Additional 1.5x boost to \feat{memory\_similarity\_score}.

- \rat{Rationale: Dense graph enables strong CF; >10 overlaps indicate similar taste.}

\textbf{Rule 2: Series Detection}

- If \feat{metadata\_overlap\_score} $> 0.8$ (likely series match): \textbf{Apply 3.0x multiplier}.

- \rat{Rationale: GoodReads users follow series religiously.}

\textbf{Rule 3: Social Signal Priority}

- If \feat{co\_interaction\_count} $> 15$:

 \;\;\;\;\;\;- Weight \feat{edge\_weight} by 0.7x (downweight pure CF).

 \;\;\;\;\;\;- Weight \feat{co\_interaction\_count} contribution by 1.5x.

- \rat{Rationale: Explicit social overlap > implicit CF in dense graphs.}

\textbf{Rule 4: Minimal Recency Decay}

- If \feat{recency\_days} $> 365$: Apply decay factor: $\exp(-0.002 \times \feat{recency\_days})$.

- \rat{Rationale: Book preferences are stable; old data remains valuable.}

\textbf{Rule 5: Memory Amplification}

- Boost by $1.0 + 1.8 \times \feat{memory\_similarity\_score}$ when \feat{co\_interaction\_count} $> 10$.

- \rat{Rationale: Memory + social signal = very strong preference indicator.}
\end{promptbox}
\caption{LLM-generated curation rules for Books and GoodReads datasets.}
\label{fig:llm_generated_curation_rules_1}
\end{figure*}

\clearpage
\begin{figure*}[h!]
\begin{promptbox}[Generated Rules: MovieTV (Recency-Critical)]
\textbf{Rule 1: Strong Recency Decay}

- If \feat{recency\_days} $> 60$: Apply decay factor: $\exp(-0.018 \times \feat{recency\_days})$.

- If \feat{recency\_days} $> 180$: Apply stronger decay: $\exp(-0.025 \times \feat{recency\_days})$.

- \rat{Rationale: Movie/TV preferences are volatile; old data loses relevance quickly.}

\textbf{Rule 2: Metadata Compensation}

- If \feat{co\_interaction\_count} $< 3$ (sparse signal): \textbf{Apply 2.8x multiplier} to \feat{metadata\_overlap\_score}.

- \rat{Rationale: Compensate for sparse CF with genre/cast similarity.}

\textbf{Rule 3: Rare CF Signal Boost}

- If \feat{co\_interaction\_count} $\ge 3$ (rare but strong): \textbf{Apply 2.5x multiplier} to \feat{edge\_weight}.

- Additional 1.8x boost to \feat{memory\_similarity\_score}.

- \rat{Rationale: Any overlap is meaningful in sparse graphs.}

\textbf{Rule 4: Memory-Guided Ranking}

- Boost by $1.0 + 1.5 \times \feat{memory\_similarity\_score}$.

- Additional 0.5x if \feat{metadata\_overlap\_score} $> 0.6$.

- \rat{Rationale: Memory captures evolving tastes better than old interactions.}

\textbf{Rule 5: Recency Threshold Filter}

- If \feat{recency\_days} $> 365$: Apply 0.3x penalty (very outdated).

- \rat{Rationale: Movies >1 year old rarely relevant unless classics.}
\end{promptbox}

\begin{promptbox}[Generated Rules: Yelp (Category-Driven)]
\textbf{Rule 1: Categorical Dominance}

- If \feat{metadata\_overlap\_score} $> 0.7$ (same cuisine + price range): \textbf{Apply 3.5x multiplier}.

- If \feat{metadata\_overlap\_score} $> 0.85$ (+ attribute match): \textbf{Apply 4.5x multiplier}.

- \rat{Rationale: Category match is essential; CF secondary.}

\textbf{Rule 2: Very Strong Recency Decay}

- If \feat{recency\_days} $> 90$: Apply decay factor: $\exp(-0.028 \times \feat{recency\_days})$.

- If \feat{recency\_days} $> 180$: Apply penalty: additional 0.5x multiplier.

- \rat{Rationale: Restaurants change rapidly; old reviews unreliable.}

\textbf{Rule 3: Attribute-Aware Memory}

- If \feat{metadata\_overlap\_score} includes attribute match: Boost \feat{memory\_similarity\_score} by 2.2x.

- \rat{Rationale: User memory + context (outdoor seating, etc.) = strong signal.}

\textbf{Rule 4: Sparse CF Handling}

- If \feat{co\_interaction\_count} $\ge 2$ (rare): \textbf{Apply 2.0x multiplier} to \feat{edge\_weight}.

- Else: Downweight CF by 0.5x, prioritize metadata.

- \rat{Rationale: Very sparse; any overlap is meaningful but metadata dominates.}

\textbf{Rule 5: Location/Price Filter}

- If \feat{metadata\_overlap\_score} $< 0.4$ (different cuisine/price): Apply 0.2x penalty (strong filter).

- \rat{Rationale: Cross-category recommendations rarely work for restaurants.}
\end{promptbox}
\caption{LLM-generated curation rules for MovieTV and Yelp datasets.}
\label{fig:llm_generated_curation_rules_2}
\end{figure*}

\clearpage

\begin{figure*}[h!]
\begin{promptbox}[Stage-R Prompt: Collaborative Memory Synthesis]
You are an intelligent memory retrieval system for personalized recommendation. Your task is to analyze the user's personal memory and collaborative memories from their neighbors to extract preference facets.

\textbf{Target User:} User \textcolor{promptkeyword}{\{user\_id\}}

\textbf{User's Personal Memory:}
\textbf{User Memory Summary:}
\textcolor{promptkeyword}{\{user\_memory\_summary\}}

\textbf{Collaborative Neighbor Memories:}
The following neighboring users and items provide collaborative signals for understanding this user's preferences:

\textbf{Collaborative Neighbors:}
\textcolor{promptkeyword}{\{formatted\_neighbor\_list\}}

\textbf{Context (Candidate Items):}
\textbf{Candidates to Rank:}
\textcolor{promptkeyword}{\{formatted\_candidate\_list\}}
(Note: These candidates are for context only, do not score them)

\textbf{Your Task:}
Analyze the user's personal memory and the collaborative memories from neighboring users and items to identify \textcolor{promptkeyword}{\{n\_facets\}} distinct preference facets that characterize this user's interests and tastes.
For each preference facet, provide:

1. A concise natural language description of the preference (e.g., "interest in mystery novels with strong female protagonists")

2. A confidence score between 0 and 1 indicating how strongly this facet is supported by the evidence

3. A list of supporting neighbors (user IDs or item IDs) that provide evidence for this facet

Additionally, identify the collaborative edges between neighboring users/items and the target user, with edge weights (0-1) indicating the strength of collaborative signal.

\textbf{Expected Output Format:}
Your response should be a JSON object with two fields:

- "facets": An array of facet objects, each containing:

  \;\;\;* "facet": A string describing the preference
  
  \;\;\;* "confidence": A number between 0 and 1
  
  \;\;\;* "supporting\_neighbors": An array of neighbor IDs (e.g., ["User-123", "Item-456"])
  
- "support\_edges": An array of edge objects, each containing:

 \;\;\;* "from": The source neighbor ID (string)
  
  \;\;\;* "to": The target user ID (string)
  
  \;\;\;* "w": The edge weight between 0 and 1 (number)
\end{promptbox}
\caption{The prompt used by $\text{LM}_{\text{Mem}}$ to synthesize high-level memory facets from retrieved collaborative neighbors in Stage-R. \textbf{Candidate items act as context to guide task-relevant synthesis.}}
\label{fig:prompt_synthesis}
\end{figure*}

\clearpage

\begin{figure*}[h!]
\begin{promptbox}[Stage-ReRank Prompt (MemRec Mode)]
You are an intelligent recommendation scoring system. Your task is to evaluate how well each candidate item matches the target user's preferences based on their personal memory and collaborative signals.

\textbf{Target User:} User \textcolor{promptkeyword}{\{user\_id\}}

\textbf{User's Current Request:}
\textcolor{promptkeyword}{\{instruction\}}

\textbf{User Preferences (Extracted from Collaborative Memories):}
Based on collaborative signals from neighboring users and items, we have identified the following preference patterns:
\textcolor{promptkeyword}{\{formatted\_facets\}}

\textbf{Candidate Item Memories:}
\textcolor{promptkeyword}{\{formatted\_item\_memories\}}

\textbf{Your Task:}
For each of the candidate items listed above, provide a relevance score between 0 and 1 that indicates how well the item matches the user's preferences:

  • 1.0 = Excellent match, highly aligned with user's facets and memory
  
  • 0.5 = Moderate match, partially relevant
  
  • 0.0 = Poor match, not aligned with user's interests

For each item, provide a brief rationale explaining your scoring decision based on the user's preference facets and personal memory.

\textbf{Expected Output Format:}
Your response should be a JSON object with a single field:

- "scores": An array of scoring objects, each containing:

  \;\;\;* "item\_id": The item's ID (integer)
  
  \;\;\;* "score": Your relevance score between 0 and 1 (number)
  
  \;\;\;* "rationale": A brief explanation of your scoring (string)
\end{promptbox}
\caption{The prompts used by $\text{LLM}_{\text{Rec}}$ for candidate scoring in Stage-ReRank.
}
\label{fig:prompt_rerank}
\end{figure*}

\clearpage
\begin{figure*}[h!]
\begin{promptbox}[Stage-W Prompt: Asynchronous Collaborative Propagation]
You are an intelligent memory management system for collaborative recommendation. Your task is to update the personal memories of the user, the clicked item, and relevant collaborative neighbors based on this new interaction.

\textbf{Interaction Context:}
User \textcolor{promptkeyword}{\{user\_id\}} has just interacted with (clicked) Item \textcolor{promptkeyword}{\{item\_id\}} (\textcolor{promptkeyword}{\{clicked\_item\_info\}}).

\textbf{User Preferences (Extracted from Collaborative Memories):}
The following preference patterns were identified for this user:
\textcolor{promptkeyword}{\{formatted\_facets\}}

\textbf{Current Personal Memory of User \{user\_id\}:}
\textcolor{promptkeyword}{\{current\_user\_memory\}}

\textbf{Current Memory of Item \textcolor{promptkeyword}{\{item\_id\}} (\textcolor{promptkeyword}{\{clicked\_item\_info\}}):}
\textcolor{promptkeyword}{\{current\_item\_memory\}}

\textbf{Collaborative Neighbors Available for Memory Propagation:}
The following \textcolor{promptkeyword}{\{n\_neighbors\}} collaborative neighbors are available for potential memory updates:
\textcolor{promptkeyword}{\{formatted\_neighbors\}}

\textbf{Your Task:}
Generate UPDATED memories for:

1. \textbf{The current user} (synthesize current memory + facets + clicked item)

2. \textbf{The clicked item} (describe what it is and who might enjoy it)

3. \textbf{Selected neighbors} (IMPORTANT: collaborative propagation is key!)

   \;\;\;* Analyze the available neighbors and their current memories
   
   \;\;\;* Select neighbors that are RELEVANT to this interaction (e.g., similar themes, related topics)
   
   \;\;\;* Update their memories to reflect new insights from this interaction
   
   \;\;\;* This helps the system learn collaboratively!

\textbf{Output Requirements:}

- "user\_memory": Concise natural language description of user's interests and preferences

  \;\;\;* Synthesize themes (e.g., "holistic health", "children's education")
  
  \;\;\;* Be specific (e.g., "interested in Reiki and aromatherapy")
  
  \;\;\;* DON'T just list item titles
  
  \;\;\;* Keep it focused (typically a few sentences)
  
- "item\_memory": Concise description of the clicked item

  \;\;\;* What it's about and who might enjoy it
  
  \;\;\;* Keep it brief but informative
  
- "neighbor\_updates": Array of neighbor memory updates (OPTIONAL but recommended)

  \;\;\;* Select neighbors that are MOST relevant to this interaction
  
  \;\;\;* Choose as many as needed (typically a few, but flexible)
  
  \;\;\;* For each neighbor, provide updated memory content (NOT just appending text)
  
  \;\;\;* Rationale explains why this neighbor is relevant

\textbf{Expected Output Format:}
Your response should be a JSON object with three fields:

- "user\_memory": The updated personal memory for the user (string)

- "item\_memory": The updated memory for the clicked item (string)  

- "neighbor\_updates": An array of neighbor update objects (may be empty), each containing:

  \;\;\;* "neighbor\_id": The neighbor's ID, e.g., "User-123" or "Item-456" (string)
  
  \;\;\;* "memory\_update": The updated memory content for this neighbor (string)
  
  \;\;\;* "rationale": A brief explanation of why this neighbor should be updated (string)
\end{promptbox}
\caption{The prompt used by $\text{LM}_{\text{Mem}}$ to asynchronously update user and neighbor memories in Stage-W.}
\label{fig:prompt_propagation}
\end{figure*}

\clearpage

\begin{figure*}[h!]
\begin{promptbox}[GPT-4o Judge Prompt for Rationale Quality Evaluation]
\textbf{SYSTEM PROMPT:}

You are an expert evaluator for recommender systems. Your task is to assess the quality of explanations (rationales) generated by three different AI agents designed to recommend items to users.

You will be provided with:

1. A summary of the target User's interests.

2. The recommended Item name.

3. Rationale A (generated by Model A).

4. Rationale B (generated by Model B).

5. Rationale C (generated by Model C).

You must evaluate each rationale independently on three distinct criteria using a 1-5 Likert scale.

\textbf{Evaluation Criteria \& Scoring Rubric:}

\textbf{1. Specificity (1-5 Points)}
Measure how concrete and detailed the rationale is regarding the recommended item.

- 1 (Vague): Very generic; could apply to many items in the category (e.g., "It's a good book").

- 3 (Moderate): Mentions general themes or genre traits but lacks specific details.

- 5 (Highly Specific): Richly detailed; mentions specific plot elements, character traits, writing style, or unique features of the item.

\textbf{2. Relevance (1-5 Points)}
Measure how well the rationale explains *why* this item suits this specific user based on their profile.

- 1 (Irrelevant): A generic recommendation unrelated to the user's known interests.

- 3 (Acceptable): Makes a basic connection to user genre preferences.

- 5 (Highly Personalized): explicitly ties specific item features to specific aspects of the user's history or tastes.

\textbf{3. Factuality (1-5 Points)}
Measure the accuracy of the claims made about the item.

- 1 (Hallucinated): Contains major factual errors or describes a different item entirely.

- 5 (Accurate): All claims about the item's content and characteristics are factually correct.

\textbf{Output Format:}
You must output strictly valid JSON immediately, without any additional text. The format should be:
\{

  \;\;\;"model\_a": \{"specificity": <int>, "relevance": <int>, "factuality": <int>\},
  
  \;\;\;"model\_b": \{"specificity": <int>, "relevance": <int>, "factuality": <int>\},
  
  \;\;\;"model\_c": \{"specificity": <int>, "relevance": <int>, "factuality": <int>\}
  
\}

\noindent\rule{\textwidth}{0.4pt}

\textbf{USER PROMPT TEMPLATE:}

User Interests Summary: \textcolor{promptkeyword}{\{user\_history\_summary\}}

Recommended Item: \textcolor{promptkeyword}{\{item\_title\}}

Rationale A: \textcolor{promptkeyword}{\{rationale\_model\_a\}}

Rationale B: \textcolor{promptkeyword}{\{rationale\_model\_b\}}

Rationale C: \textcolor{promptkeyword}{\{rationale\_model\_c\}}

Please evaluate Rationale A, Rationale B, and Rationale C based on the system instructions and provide the JSON output.
\end{promptbox}
\caption{The system prompt and user input template used for the GPT-4o based rationale quality evaluation. The judge evaluates three models simultaneously across Specificity, Relevance, and Factuality domains.}
\label{fig:app_prompt_rationale_eval}
\end{figure*}